\documentclass[aps,pre,preprintnumbers,showpacs,superscriptaddress]{revtex4}

\usepackage{amsmath}
\usepackage{graphicx}
\usepackage{epstopdf}

\newcommand{\erf}{{\rm erf}}

\begin{document}

\title{Extreme Fluctuations in Stochastic Network Coordination with Time Delays}

\author{D. Hunt\footnote{present address: Department of Biomathematics, David Geffen School of Medicine at UCLA, Los Angeles, CA 90095, USA}}
\affiliation{Department of Physics, Applied Physics, and Astronomy}
\affiliation{Network Science and Technology Center}
%\affiliation{Social and Cognitive Networks Academic Research Center}

\author{F. Moln\'ar, Jr.\footnote{present address: Department of Physics and Astronomy, Northwestern University, 2145 Sheridan Rd., Evanston, IL 60208, USA}}
\affiliation{Department of Physics, Applied Physics, and Astronomy}
\affiliation{Network Science and Technology Center}
%\affiliation{Social and Cognitive Networks Academic Research Center}

\author{B.K. Szymanski}
\affiliation{Department of Computer Science \\
Rensselaer Polytechnic Institute, 110 8$^{th}$ Street, Troy, NY 12180--3590, USA}
\affiliation{Network Science and Technology Center}
%\affiliation{Social and Cognitive Networks Academic Research Center}

\author{G. Korniss\footnote{Corresponding author. korniss@rpi.edu}}
%\author{G. Korniss}
%\email{korniss@rpi.edu}
\affiliation{Department of Physics, Applied Physics, and Astronomy}
\affiliation{Network Science and Technology Center}
%\affiliation{Social and Cognitive Networks Academic Research Center}

%%%%%%%%%%%%%%%%%%%%%%%%%%%%%%%%

\begin{abstract}
We study the effects of uniform time delays on the extreme
fluctuations in stochastic synchronization and coordination problems
with linear couplings in complex networks. We obtain the average
size of the fluctuations at the nodes from the behavior of the
underlying modes of the network. We then obtain the scaling behavior
of the extreme fluctuations with system size, as well as the
distribution of the extremes on complex networks, and compare them
to those on regular one-dimensional lattices. For large complex
networks, when the delay is not too close to the critical one,
fluctuations at the nodes effectively decouple, and the limit
distributions converge to the Fisher-Tippett-Gumbel density. In
contrast, fluctuations in low-dimensional spatial graphs are
strongly correlated, and the limit distribution of the extremes is
the Airy density.
%%%%%%%%%%%%%%%%%%%%%%%%%%%%%%%%%%%%%%%%%%%%%%%%%%%%%%%%%%%%%%%%%%%
Finally, we also explore the effects of nonlinear couplings on the stability and on the extremes of the synchronization landscapes.
\end{abstract}


\date{\today}
\maketitle

\section{Introduction}

Synchronization and coordination involve a system of coupled,
autonomously interacting units or agents attempting to achieve a
common goal \cite{Saber_IEEE2007, Arenas_PhysRep2008,
Korniss_Springer2012, Sipahi_IEEE2011}. Synchronization of a system
emerges from the cumulative efforts of the individual entities, each
regulating themselves based on the information they can gather from
their neighbors on the system's local state. The difficulties in
synchronization or coordination problems are often compounded by
stochastic effects and time delays
\cite{HuntPRL,HuntPLA,HuntPRE,Hod_PRL2010,Chen_EPL2008,Chen_PRE2009,Chen_PLOS2011},
preventing global coordination or consensus. Time delays between the
state of the system and the reaction to that information (due to
e.g. transmission, cognition, or execution) can pose
significant challenges. Critical aspects of the underlying theory of
delays have been long established in the context of macro-economic
cycles as far back as 1935 \cite{Kalecki_1935, Frisch_1935}. In such
cases, the description of the complex system can be reduced to a
single stochastic variable \cite{Kuechler_SSR1992, Ohira_PRE2000,
Frank_PRE2001}. Recent interest in the application of time delays to
complex networks \cite{Saber_IEEE2007, Huberman_IEEE1991,
Strogatz_PRE2003} provides fresh insights extending these older
results. Understanding the dynamics across a complex network offers
the possibility to optimize synchronization
\cite{Arenas_PhysRep2008, Barahona_PRL2002, Nishikawa_PRL2002,
LaRocca_PRE2008, LaRocca_PRE2009}, including weighted graphs
\cite{Zhou_PRL2006, GK_PRE2007, Korniss_Springer2012,Lai_Chaos2008}.
Synchronization and coordination with delays has been studied in the
stock market \cite{Saavedra_PNAS2012}, ecological systems
\cite{fireflies, Cucker_IEEE2007, Vicsek_PRL1995, Reynolds_CG1987},
population dynamics \cite{Hutchinson_1948, May_1973,Ruan_2006}, postural sway
and balance \cite{Milton_EPL2008, Milton_PTRSA2009, Cabrera_PRL2002,
Cabrera_CMP2006}, and the human brain \cite{Izhikevich_SIAM2001,
Chen_EPL2008, Chen_PRE2009, Chen_PLOS2011}. It is also important to
understand critical functions of autonomous artificial systems, such
as congestion control in networks \cite{Saber_IEEE2007, GK_PRE2007,
Korniss_Springer2012, Johari_IEEE2001, Saber_IEEE2004, Ott_2006},
massively parallel \cite{GK_Science2003, Korniss_PRL2000} and
distributed computing \cite{Guclu_PRE2006, Guclu_Chaos2007}, and
vehicular traffic \cite{Orosz_PRSA2006, Orosz_PTRSA2010,
Fax_IEEE2004, Saber_IEEE2004}. The aim of this paper is to explore
the effects of noise and delays on the dynamics in complex and
random networks
\cite{Watts_Nature1998,Barab_sci,BarabREV,MendesREV,ER_1960},
specifically on the extreme fluctuations.
%%%%%%%%%%%%%%%%%%%%%%%%%%%%%%%%%%%%%%%%%%%%%%%%%%%%%%%%%%%%%%%%%%%
Extreme fluctuations can have critical implications in synchronization, coordination, or load balancing problems,
since large-scale or global system failures are often triggered by extreme events occurring on an individual
node \cite{Guclu2004,Guclu_FNL,Lai_SREP2014}.
%%%%%%%%%%%%%%%%%%%%%%%%%%%%%%%%%%%%%%%%%%%%%%%%%%%%%%%%%%%%%%%%%%%%
In order to show the implications of the general theoretical results, we will cover the
implications for typologically distinct networks.

The scaling behavior of extreme fluctuations in the case of zero
time delay has been investigated previously for small-world (SW)
\cite{Guclu2004,Guclu_FNL} and scale-free (SF) networks
\cite{Guclu_Chaos2007}, as well as low-dimensional regular
topologies \cite{Guclu_Chaos2007}. Despite having more complex
interaction topologies, coordination and synchronization phenomena
of the former systems (as far as critical behavior is concerned)
actually tend to be simpler than those of their low-dimensional
regular-topology counterparts. This is because fluctuations of the
relevant field-variables at the nodes are weakly correlated in
complex networks \cite{Guclu_FNL,Guclu2004,Guclu_Chaos2007}. Hence,
standard extreme-value limit theorems apply to the statistics of the
extremes (as well as to those of the system-averaged fluctuations,
i.e., the width) \cite{Guclu_FNL}. In contrast, fluctuations in
one-dimensional regular lattices are strongly correlated, and the
applicability of traditional extreme-value limit theorems
immediately break down
\cite{Guclu_Chaos2007,Majumdar_2004,Majumdar_2005,Raychaudhuri_PRL2001}
(as well as limit theorems for the sum of local variables
\cite{FORWZ_PRE1994}).

While extreme-value theory for the scaling properties and universal
limit-distributions of uncorrelated (or weakly-correlated) random
variables is well established \cite{Fisher1928,Gumbel1958,
Galambos1994}, only a few results are available on statistical
properties of the extremes of strongly correlated variables
\cite{Majumdar_2004,Majumdar_2005}. Majumdar and Comtet obtained the
distribution of extreme fluctuations in a correlated stochastic
one-dimensional landscape only recently \cite{Majumdar_2004} (with
no time delays). In coupled interacting systems with no delays,
possible divergences of the width and the extremes are associated
with the small-eigenvalue behavior of the Laplacian spectrum (e.g.,
with long-wavelength modes in low-dimensional systems or low
connectivity in complex networks)
\cite{Guclu_Chaos2007,Guclu_FNL,Majumdar_2004,Majumdar_2005,Raychaudhuri_PRL2001}.
In the presence of time delays, however, singularities and
instabilities can also be governed by the largest eigenvalues when
the system is close to the synchronizability threshold
\cite{HuntPRL,HuntPRE}. To that end, we investigate finite-size
effects and the universality class of the extreme fluctuations in
complex networks stressed by time delays.

In parallel to the sum of a large number of uncorrelated (or
weakly-correlated) {\em short-tailed} random variables approaching a
Gaussian distribution (governed by the central-limit theorem), the
largest of these (suitably-scaled) variables converges to the
Fisher-Tippett-Gumbel (FTG) \cite{Fisher1928, Gumbel1958,
Galambos1994} (cumulative) distribution,
%%%%%%%%%%%%%%%%%%%%%%%%%%%%%%%%%%%%%%%%%%%%%%%%%%%%%%%%%%%%%%%%%%%
\begin{equation}
P_{\max}^{<}(\tilde x) \simeq e^{-e^{-\tilde x}}\;,
\end{equation}
%%%%%%%%%%%%%%%%%%%%%%%%%%%%%%%%%%%%%%%%%%%%%%%%%%%%%%%%%%%%%%%%%%%
where $\tilde{x} = (x_{\max} - a_N)/b_N$ is the scaled extreme
\cite{exp_tail}, with mean $\langle\tilde{x}\rangle=\gamma$
($\gamma=0.577\ldots$ being the Euler constant) and variance
$\sigma_{\tilde{x}}^2 = \langle\tilde{x}^2\rangle -
\langle\tilde{x}\rangle^2=\pi^2/6$. The expected largest value of
the original variables (e.g., for exponential-like tails
\cite{exp_tail}) scales as
%%%%%%%%%%%%%%%%%%%%%%%%%%%%%%%%%%%%%%%%%%%%%%%%%%%%%%%%%%%%%%%%%%%
\begin{equation}
\langle x_{\max}\rangle = a_N + b_N\gamma \simeq \left(\frac{\ln N}{c}\right)^{1/\delta}.
\label{x_max}
\end{equation}
%%%%%%%%%%%%%%%%%%%%%%%%%%%%%%%%%%%%%%%%%%%%%%%%%%%%%%%%%%%%%%%%%%%
Note that corrections to this scaling are of ${\cal O}(1/\ln N)$,
which can be noticeable in finite-size networks that are
computationally feasible in our investigations.
%%%%%%%%%%%%%%%%%%%%%%%%%%%%%%%%%%%%%%%%%%%%%%%%%%%%%%%%%%%%%%%%%%
For comparison with numerical data, it is often convenient to employ
the extreme-value limit distribution of the variable scaled to zero
mean and unit variance, $y=(x_{\max}-\langle
x_{\max}\rangle)/\sigma_{x_{\max}}$,
%%%%%%%%%%%%%%%%%%%%%%%%%%%%%%%%%%%%%%%%%%%%%%%%%%%%%%%%%%%%%%%%%%%
\begin{equation}
P_{\max}^{<}(y) = e^{-e^{-(ay+\gamma)}} \;,
\end{equation}
%%%%%%%%%%%%%%%%%%%%%%%%%%%%%%%%%%%%%%%%%%%%%%%%%%%%%%%%%%%%%%%%%%%
where $a=\pi/\sqrt{6}$. The corresponding FTG density then becomes
%%%%%%%%%%%%%%%%%%%%%%%%%%%%%%%%%%%%%%%%%%%%%%%%%%%%%%%%%%%%%%%%%%%
\begin{equation}
p_{\max}^{<}(y) = a e^{-(ay+\gamma)-e^{-(ay+\gamma)}} \;.
\label{FTG_density}
\end{equation}
%%%%%%%%%%%%%%%%%%%%%%%%%%%%%%%%%%%%%%%%%%%%%%%%%%%%%%%%%%%%%%%%%%%

We hypothesize that the FTG limit distributions of the extreme
fluctuations in stochastic network synchronization will also be
applicable to the case of nonzero time delays, provided that the
large but finite system is in the synchronizable regime. Although
the fluctuations at the nodes will, of course, depend on the delay,
the system can be considered as a collection of a large number of
weakly-correlated components. In contrast, in the case of a
one-dimensional regular lattice (ring) with delayed coupling,
we expect that the limit distribution of the extreme fluctuations approaches the Airy
distribution \cite{Guclu_Chaos2007,Majumdar_2004,Majumdar_2005,OMalley_2008}.

Thus, provided that the system is synchronizable, the scaling with
the system size and the shape and class of the respective
extreme-value limit distributions will be the same as those of a
network without time delays. To put it simply, time delays will
impact the ``prefactors" (within the syncronizable regime), but not
the extreme-value universality class. The focus of this paper is to
test the above hypotheses.

\section{Eigenmode Decomposition, Fluctuations, and the Width}

In the simplest linear synchronization or coordination problem in
networks with delay, the relevant (scalar) variable at each node
evolves according to
\begin{equation}
\partial_t h_i(t) = -\sum_{j = 1}^N C_{ij}[h_i(t - \tau) - h_j(t - \tau)] + \eta_i(t) = -\sum_{j = 1}^N \Gamma_{ij} h_j(t - \tau) + \eta_i(t) \;,
\label{nodeDiffEq}
\end{equation}
where $C_{ij}$ is the coupling matrix and $\Gamma_{ij} = -C_{ij} +
\delta_{ij}\sum_\ell C_{i\ell}$ is the (symmetric) network
Laplacian. Here, we consider unweighted graphs, thus $C_{ij}$ is
just the adjacency matrix and $\sum_\ell C_{i\ell}=k_i$ is the
degree of node $i$. The noise $\eta_i(t)$ is Gaussian with
zero mean and correlations $\langle \eta_i(t)\eta_j(t')\rangle =
2D\delta_{ij}\delta(t - t')\rangle$. In our simulations, without
loss of generality, we set $D=1$. We have previously studied the
behavior of the average size of the fluctuations about the mean for
a network with noise and time delays \cite{HuntPRL,HuntPRE}, i.e.,
the width
%%%%%%%%%%%%%%%%%%%%%%%%%%%%%%%%%%%%%%%%%%%%%%%%%%%%%%%%%%%%%%%%%%%%
\begin{equation}
\langle w^2(t)\rangle
    = \left\langle \frac 1 N \sum_{i = 1}^N[h_i(t) - \bar h(t)]^2\right\rangle \;,
\label{theWidth}
\end{equation}
%%%%%%%%%%%%%%%%%%%%%%%%%%%%%%%%%%%%%%%%%%%%%%%%%%%%%%%%%%%%%%%%%%%%
where $\bar h_(t) = N^{-1}\sum_i h_i(t)$ is the mean at time $t$
and $\langle\cdot\rangle$ indicates averaging over different
realizations of the noise. In the present paper, we are interested
in the extremes of the fluctuations in the system at a given time.
Because of the symmetry about the mean of the relaxation term in Eq.
(\ref{nodeDiffEq}), the distribution of extreme fluctuations above
and below the mean are identical, so we will reduce the presentation
of results to those of the maximum of a snapshot, given by
\begin{equation}
\Delta_{\max}(t)
    = \max_i\{\Delta_i(t)\}
\end{equation}
where $\Delta_i(t) = h_i(t) - \bar h(t)$ is the fluctuation about the mean of an individual node.

\begin{figure}[t]
\vspace{1.0truecm}
\centering
\includegraphics[scale=0.6]{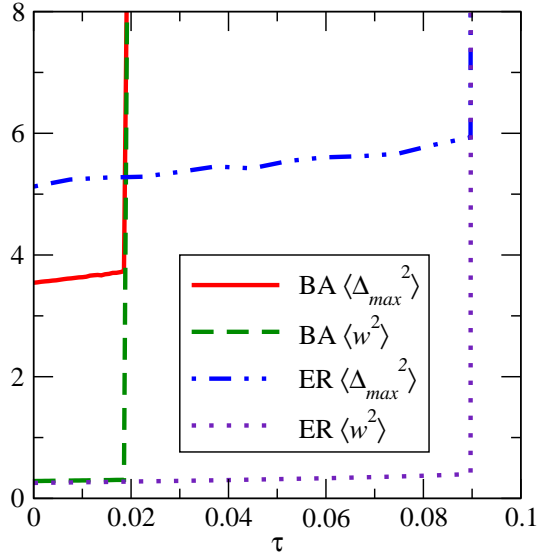}
\caption{(Color online) The typical behavior of the steady-state average width and
the expected extreme as a function of the delay $\tau$ for an ER and a BA
network with $N$$=$$1000$ and $\langle k\rangle \approx 6$.}
\label{scalingExtremeAndAverage}
\end{figure}

Diagonalizing $\Gamma$ from Eq.~(\ref{nodeDiffEq}) gives $N$ independent modes $\tilde h_k(t)$, each of which obey an equation of the form
\begin{equation}
\partial_t\tilde h_k
    = -\lambda_k \tilde h_k(t - \tau) + \tilde\eta_k(t) \;,
\label{hk_sde}
\end{equation}
where $\lambda_k$ is the corresponding eigenvalue for mode $k$.
Organizing the labels of the modes such that $0\leq\lambda_k \le
\lambda_{k + 1}$, a network with positive, symmetric couplings and a
single connected component has a single (and uniform) mode
associated with $\lambda_0 = 0$, which does not contribute to
fluctuations about the mean and so does not impact either the width
or the extremes, as both are measured from the mean. The condition
for the average fluctuations $\langle\tilde h^{2}_k\rangle$ to
remain finite in the steady-state for Eq.~(\ref{hk_sde}) is known
exactly
\cite{Kuechler_SSR1992,Saber_IEEE2004,Frisch_1935,Hayes_1950,HuntPRL},
%%%%%%%%%%%%%%%%%%%%%%%%%%%%%%%%%%%%%%%%%
\begin{equation}
\lambda_k\tau < \pi/2 \;.
\end{equation}
%%%%%%%%%%%%%%%%%%%%%%%%%%%%%%%%%%%%%%%%%
Hence, for the network to remain synchronizable, the above must hold
for all $k>0$, or equivalently,
%%%%%%%%%%%%%%%%%%%%%%%%%%%%%%%%%%%%%%%%%
\begin{equation}
\tau < \tau_c \equiv \pi/2\lambda_{\max} \;.
\label{tau_c}
\end{equation}
%%%%%%%%%%%%%%%%%%%%%%%%%%%%%%%%%%%%%%%%%
This condition guarantees that the system avoids delay-induced
instabilities and that both the width and the extremes will have a
finite steady-state value. Further, for the simple stochastic
differential equation with delay in Eq.~(\ref{hk_sde}), the
steady-state variance of the corresponding stochastic variable is
also known exactly \cite{Kuechler_SSR1992},
%%%%%%%%%%%%%%%%%%%%%%%%%%%%%%%%%%%%%%%%%%%%%%%%%%%%%%%%%%%%%%%%%%%%%
\begin{equation}
\langle\tilde{h}^{2}_{k}\rangle\ = D\tau f(\lambda_k\tau)
\equiv
D\tau \frac{1 +
\sin(\lambda_k\tau)}{\lambda_k\tau\cos(\lambda_k\tau)}
=
D \frac{1 +
\sin(\lambda_k\tau)}{\lambda_k\cos(\lambda_k\tau)}\;.
\label{f_scaling}
\end{equation}
%%%%%%%%%%%%%%%%%%%%%%%%%%%%%%%%%%%%%%%%%%%%%%%%%%%%%%%%%%%%%%%%%%%%%
Hence, given the eigenvalues of the Laplacian for a given network,
one has an exact expression for the average steady-state width as
well \cite{HuntPRL,HuntPRE},
%%%%%%%%%%%%%%%%%%%%%%%%%%%%%%%%%%%%%%%%%%%%%%%%%%%%%%%%%%%%%%%%%%%%%
\begin{equation}
\langle w^2 \rangle
= \frac{1}{N} \sum_{i = 1}^{N} \langle\Delta^{2}_{i}\rangle
= \frac{1}{N} \sum_{k = 1}^{N - 1} \langle\tilde{h}^{2}_{k}\rangle
= \frac{D\tau}{N}\sum_{k = 1}^{N - 1} f(\lambda_k\tau)\;.
\label{exactWidth}
\end{equation}
%%%%%%%%%%%%%%%%%%%%%%%%%%%%%%%%%%%%%%%%%%%%%%%%%%%%%%%%%%%%%%%%%%%%%%
Of course, for a typical large complex network one does not have
the eigenvalues explicitly in hand. Nevertheless, one can obtain them through
numerical diagonalization. Hence, employing Eq.~(\ref{exactWidth})
provides an alternative to direct simulations of the coupled
stochastic differential equations with delay Eq.~(\ref{nodeDiffEq}).
Equation~(\ref{exactWidth}), after Taylor expansion of $\tau
f(\lambda_k\tau)$ in the variable $\tau$ in Eq.~(\ref{f_scaling}), also allows one to obtain the
approximate behavior of the steady-state width (within the synchronizable regime
$\tau<\tau_c$),
%%%%%%%%%%%%%%%%%%%%%%%%%%%%%%%%%%%%%%%%%%%%%%%%%%%%%%%%%%%%%%%%%%%%%
\begin{eqnarray}
\langle w^2 \rangle_{\tau} & = & \frac{D}{N}\sum_{k = 1}^{N - 1} \tau f(\lambda_k\tau) =
\frac{D}{N}\sum_{k = 1}^{N - 1} \frac{1}{\lambda_k} +
\frac{D}{N}\sum_{k = 1}^{N - 1} \tau +
\frac{D}{N}\sum_{k = 1}^{N - 1} \frac{\tau^2}{2}\lambda_k + {\cal O}(\tau^3) \nonumber \\
& = &
\langle w^2 \rangle _{\tau=0} +
D\frac{N-1}{N}\tau +
\frac{D\langle k\rangle}{2}\tau^2  + {\cal O}(\tau^3)
\simeq
\langle w^2 \rangle _{\tau=0} +
D\tau +
\frac{D\langle k\rangle}{2}\tau^2 \;,
\label{w2_tau}
\end{eqnarray}
%%%%%%%%%%%%%%%%%%%%%%%%%%%%%%%%%%%%%%%%%%%%%%%%%%%%%%%%%%%%%%%%%%%%%%
for large networks ($1/N$$\ll$$1$). In obtaining the above expression we exploited
that the trace is invariant under basis transformation, hence,
%%%%%%%%%%%%%%%%%%%%%%%%%%%%%%%%%%%%%%%%%%%%%%%%%%%%%%%%%%%%%%%%%%%%%%%%%%%%%%%%%%%%%%%%%%%%%%%%%%%%
$\sum_{k = 1}^{N - 1} \lambda_k = \sum_{i = 1}^{N} \Gamma_{ii} = \sum_{i = 1}^{N} k_i = N\langle k\rangle$
%%%%%%%%%%%%%%%%%%%%%%%%%%%%%%%%%%%%%%%%%%%%%%%%%%%%%%%%%%%%%%%%%%%%%%%%%%%%%%%%%%%%%%%%%%%%%%%%%%%%
for an unweighted graph. The first term above is the width for the
network with no delay, which depends strongly on the detailed
structure of the graph through its Laplacian spectrum, $\langle w^2
\rangle _{\tau=0} = \frac{D}{N}\sum_{k = 1}^{N - 1} \lambda_k^{-1}$
\cite{GK_PRE2007}. The first-order correction in a network with
delays is {\em completely independent} of any structural
characteristics of the network. The second-order correction only
depends on the average degree (average connectivity), but is
independent of the local connectivity of the nodes, or the specific
shape and heterogeneity of the degree distribution. The behavior of
the width as a function of the delay for a Barab\'asi-Albert (BA)
scale-free network \cite{Barab_sci} and an Erd\H{o}s-R\'enyi (ER)
\cite{ER_1960} graph is shown in
Fig.~\ref{scalingExtremeAndAverage}, indicating an approximately
linear behavior for a significant portion of the synchronizable
regime, in accordance with the above prediction
[Eq.~(\ref{w2_tau})]. For comparison, the analogous behavior of the
largest fluctuations is also shown, indicating (as expected) that
the steady-state width $\langle w^2\rangle$ and the extreme
fluctuations $\langle\Delta^2_{\max}\rangle$ will diverge at the
same critical delay [Eq.~\ref{tau_c}].
%%%%%%%%%%%%%%%%%%%%%%%%%%%%%%%%%%%%%%%%%%%%%%%%%%%%%%%%%%%%%%%%%%%%%
%%%%%%%%%%%%%%%%%%%%%%%%%%%%%%%%%%%%%%%%%%%%%%%%%%%%%%%%%%%%%%%%%%%%%%%

\section{Extreme Fluctuations}

\begin{figure}[t]
\vspace{1.0truecm}
\centering
\includegraphics[scale=0.6]{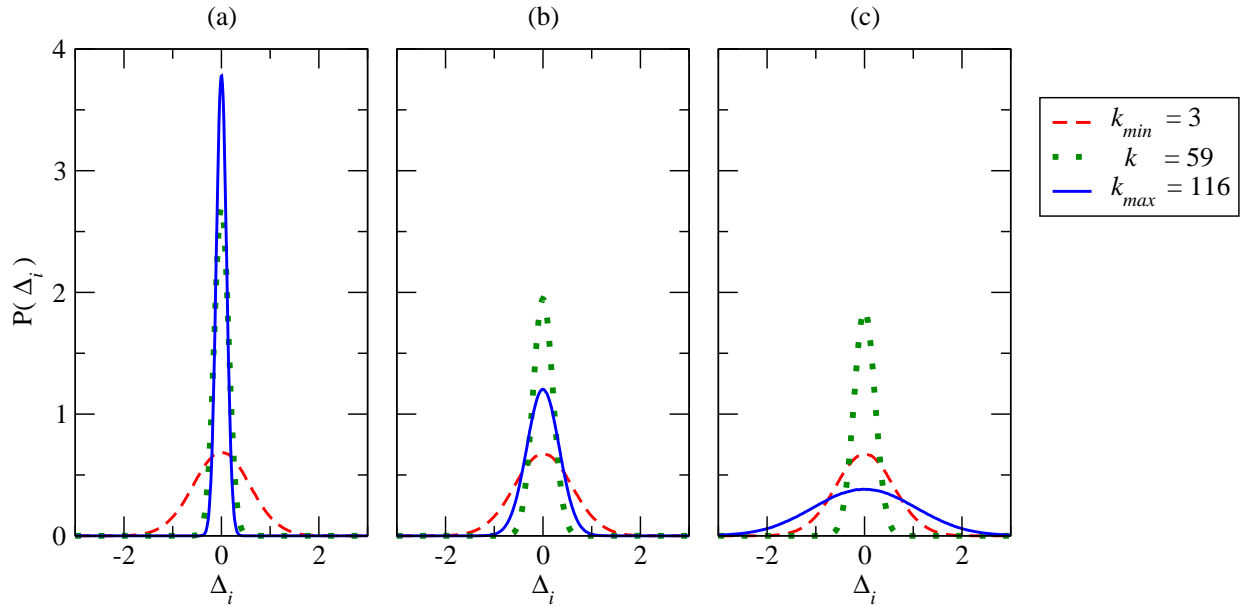}
\caption{(Color online) Individual node distributions of representative degrees in a BA network with $N$$=$$1000$ and $\langle k\rangle \approx 6$
for fractions $q$ of the critical delay of (a) $q=0$, (b) $q=0.90$, and (c) $q=0.99$ ($q$$=$$\tau/\tau_c$).
}
\label{individualNodeDistributions}
\end{figure}

With an understanding of the typical fluctuations of the underlying
modes, we may now proceed to consider the extreme fluctuations in a
network. Consider the covariance matrix of fluctuations at the nodes
(i.e., the steady-state equal-time correlations), $\sigma^2_{ij}
\equiv \langle \Delta_i\Delta_j\rangle$, and that of the modes,
%%%%%%%%%%%%%%%%%%%%%%%%%%%%%%%%%%%%%%%%%%%%%%%%%%%%%%%%%%%%%%%%%%
$\tilde\sigma^2_{k\ell} \equiv \langle \tilde{h}_{k}\tilde{h}_{\ell}
\rangle = \delta_{k\ell}\tilde\sigma^2_{k}$ (where the single
subscript denotes the diagonal elements).
%%%%%%%%%%%%%%%%%%%%%%%%%%%%%%%%%%%%%%%%%%%%%%%%%%%%%%%%%%%%%%%%%
The distribution of a single mode follows a zero-mean normal
(Gaussian) distribution with a variance given by Eq.~(\ref{f_scaling}),
\begin{equation}
\tilde \sigma_k^2
    = \langle \tilde h_k^2\rangle = D\tau f(\lambda_k\tau) \;.
\label{sigma_lambda}
\end{equation}
In turn, the fluctuations from the modes translate back to those at
the nodes according to $\boldsymbol\sigma^2 =
S\tilde{\boldsymbol\sigma}^2S^{-1}$, where $S$ is an orthogonal matrix with
columns composed of the normalized eigenvectors of the network
Laplacian (i.e., $\Gamma = S\Lambda S^{-1}$, where $\Lambda$ is a
diagonal matrix of the eigenvalues). Since $\tilde{\boldsymbol\sigma}^2$
is diagonal, this transformation can be written simply as
%%%%%%%%%%%%%%%%%%%%%%%%%%%%%%%%%%%%%%%%%%%%%%%%%%%%%%%%%%%%%%%%%%%%%%%%%%%%%%%%%%%
\begin{equation}
\langle \Delta_i^2\rangle = \sigma^2_i = \sum_k S_{ik}^2\tilde\sigma^2_k\;.
\label{sigma_i}
\end{equation}
%%%%%%%%%%%%%%%%%%%%%%%%%%%%%%%%%%%%%%%%%%%%%%%%%%%%%%%%%%%%%%%%%%%%%%%%%%%%%%%%%%%
The marginal distributions of the fluctuations at the nodes
($x=\Delta_i$) are Gaussian,
%%%%%%%%%%%%%%%%%%%%%%%%%%%%%%%%%%%%%%%%%%%%%%%%%%%%%%%%%%%%%%%%%%%%%%%%%%%%%%%%%%
\begin{equation}
p_i(x) = \frac{1}{\sqrt{2\pi\sigma^2_i}} e^{-\frac{x^2}{2\sigma_i^2}}
\label{individualNodeDistribution}
\end{equation}
with zero mean and variance $\sigma^2_i$ [Eq.~(\ref{sigma_i})].
The effects of several delays on the spread of the distributions for
a few representative degree classes are shown in
Fig.~\ref{individualNodeDistributions}.
%%%%%%%%%%%%%%%%%%%%%%%%%%%%%%%%
%%%%%%%%%%%%%%%%%%%%%%%%%%%%%%%%
Each panel shows the distributions for a distinct delay
$\tau$, which can be expressed in terms of the fraction
{\em relative} to the critical delay $q \equiv \tau/\tau_c$.

\begin{figure}[t]
\vspace{1.0truecm}
\centering
\includegraphics[scale=0.5]{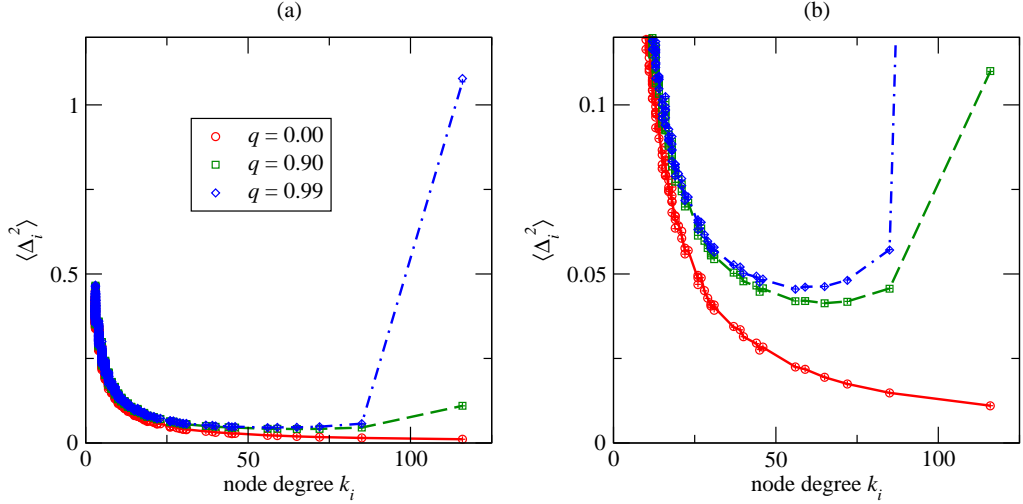}
\caption{(Color online)
(a) Average size of the fluctuations at the nodes as a function of their degree for a BA network with $N$$=$$1000$ and $\langle k\rangle$$\approx$$6$.
Open symbols correspond to results based on exact numerical diagonalization of the Laplacian and employing Eqs.~(\ref{sigma_lambda}) and (\ref{sigma_i}).
Plus symbols (of matching colors) correspond to the direct numerical integration of the stochastic delay-differential Eq.~(\ref{nodeDiffEq})
with $\Delta t$$=$$5\times 10^{-6}$. The connecting lines are the average of the degree class from these numerical integrations.
(b) Same data as in (a) but on smaller vertical scales.}
\label{individualNodeVariances}
\end{figure}

For zero or small delays the size of the fluctuations (the width of
the distributions) at a node decreases monotonically with the node's
degree, i.e., the larger the degree the narrower the distribution
[Fig.~\ref{individualNodeDistributions}(a,b)]. For a sufficiently
large delay, however, the trend changes, and the node with the
largest degree can exhibit the largest fluctuations
[Fig.~\ref{individualNodeDistributions}(c)]. This can be understood
by noting that in a mean-field sense, the {\em effective coupling}
at each node is its degree $k_i$
\cite{Korniss_Springer2012,GK_PRE2007}. Thus, the fluctuations at
each node are approximately proportional to $f(k_i\tau)$ (where the
scaling function is known exactly [Eq.~(\ref{f_scaling})]), and it
is {\em non-monotonic} in its argument \cite{HuntPRL,HuntPRE}.
%%%%%%%%%%%%%%%%%%%%%%%%%%%%%%%%%%%%%%%%%%%%%%%%%%%%%%%%%%%%%%%%%%%%%%
This trend is also illustrated in
Fig.~\ref{individualNodeVariances}. For zero (or small) delay the
average fluctuations at a node decay as
$\langle\Delta_{i}^2\rangle\sim1/k_i$
\cite{Korniss_Springer2012,GK_PRE2007}. In contrast, the average
size of the fluctuations as a function of the degree becomes
non-monotonic for large delays [Fig.~\ref{individualNodeVariances}].
The fluctuations at the low-degree nodes remain largely unaffected,
while fluctuations at the large-degree nodes increase significantly.
%%%%%%%%%%%%%%%%%%%%%%%%%%%%%%%%%%%%%%%%%%%%%%%%%%%%%%%%%%%%%%%%%%%%%
%%%%%%%%%%%%%%%%%%%%%%%%%%%%%%%%

\begin{figure}[t]
\vspace{1.0truecm}
\centering
\includegraphics[scale=0.6]{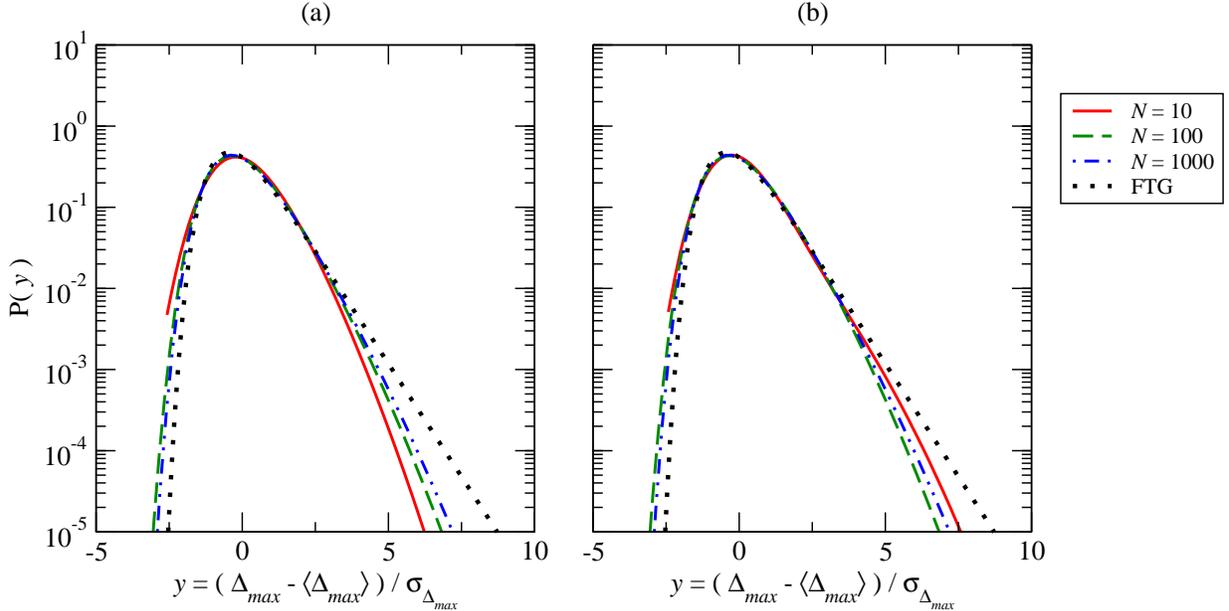}
\caption{(Color online) Finite-size behavior of the distribution of the extreme
fluctuations for BA networks with $\langle k\rangle$$\approx$$6$ and relative delays (a) $q$$=$$0.5$ and
(b) $q$$=$$0.9$ obtained by numerically integrating Eq.~(\ref{nodeDiffEq}) using $\Delta t$$=$$5 \times 10^{-6}$.
The dotted curves in both panels correspond to the scaled FTG density, Eq.~(\ref{FTG_density}).}
\label{extreme_fss}
\end{figure}

For networks with no delays it has been established that the nodes
with the smallest degree typically contribute most to the extremes
\cite{GK_PRE2007, Guclu_Chaos2007}, which is still valid in the case
of small delays ($\tau/\tau_c \ll 1$). For scale-free (SF) networks
with power-law degree distributions, such as BA networks
\cite{Barab_sci}, the low-degree nodes can still dominate the
distribution of extreme fluctuations at higher delays (but $\tau <
\tau_c$) since they are more numerous, even though the typical
fluctuations for the highest degree node are larger than for a
single low degree node. So long as the highest degree node's
fluctuations do not dominate the extremes of the network, the large
set of lowest degree nodes will lead to the familiar FTG
distribution for the network's extremes [Fig.~\ref{extreme_fss}(a)].
Note that the approach to the FTG limit distribution can be very
slow due to the slowly vanishing corrections for Gaussian-like
individual variables \cite{exp_tail}. Further, for larger delays,
the convergence to the FTG density may not be monotonic due to the
larger effect of the largest-degree node for small system sizes
[Fig.~\ref{extreme_fss}(b)].

Note that the largest eigenvalue of the network Laplacian varies
among individual realizations of a random network ensemble.
Therefore, to simulate ``similar" synchronization dynamics in a
network random ensemble (e.g., of $1000$ networks of size $N$), we
kept $q$, the fraction of the delay relative to the critical delay,
fixed in the individual network realizations (i.e., an individual
network realization $\ell$ has a delay $\tau^{(\ell)} =
q\tau_c^{(\ell)}$). Further, also note that the largest eigenvalue
of the Laplacian diverges with the largest degree in a graph
\cite{Fiedler_1973,Anderson_1985}, hence it diverges with the
network size $N$ in complex networks, e.g., in a power-law fashion
for SF networks \cite{Boguna_EPJB2004,{Catanzaro_PRE2005}}, and
logarithmically for ER and SW networks \cite{HuntPRL,HuntPRE}.
%%%%%%%%%%%%%%%%%%%%%%%%%%%%%%%%%%%%%%%%%%%%%%%%%%%%%%%%%%%%%%%%%%%%
%%%%%%%%%%%%%%%%%%%%%%%%%%%%%%%%%%%%%%%%%%%%%%%%%%%%%%%%%%%%%%%%%%%%

\begin{figure}[t]
\vspace{1.0truecm}
\centering
\includegraphics[scale=0.7]{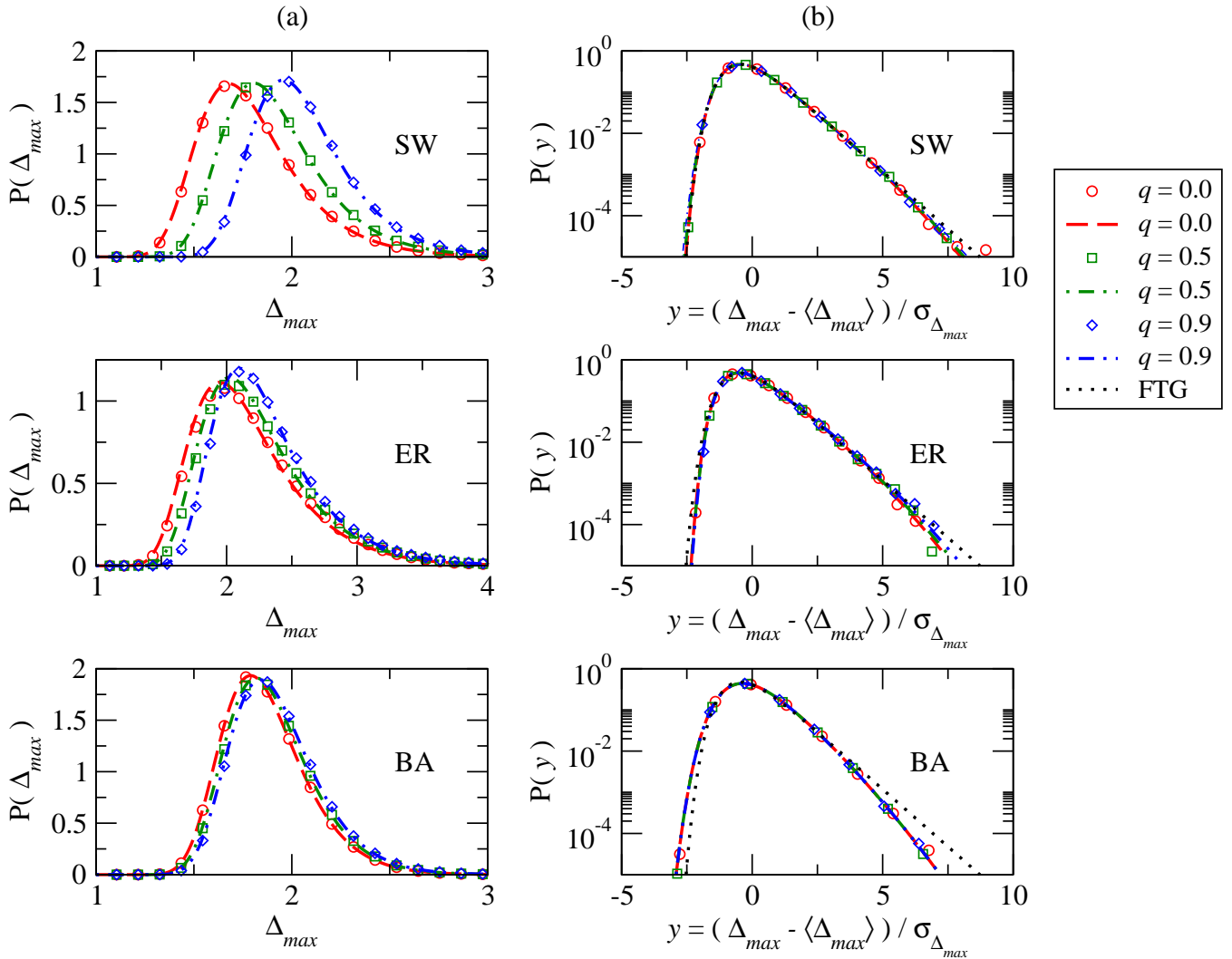}
\caption{(Color online)
(a) Extreme fluctuation distributions and (b) rescaling of
the same for several delays for SW, ER, and BA networks with $N$$=$$1000$ and $\langle k\rangle$$\approx$$6$.
The various curves correspond to the same delay fraction $q$ for both (a) and (b).
Lines are predictions based on exact numerical diagonalization and employing Eqs.~(\ref{sigma_lambda}),
(\ref{sigma_i}), and (\ref{extremeDistributionDensity}).
Symbols correspond to the numerical integration of the stochastic delay-differential equations Eq.~(\ref{nodeDiffEq})
with $\Delta t = 5 \times 10^{-6}$. The dotted curves in all panels
correspond to the scaled FTG density, Eq.~(\ref{FTG_density}).}
%(c) $\Delta_{\max}$ and $w^2$ for a BA network with $N = 1000$ and $\langle k\rangle \approx 6$.}
\label{extremeDistributions}
\end{figure}

\begin{figure}[t]
\vspace{1.0truecm}
\centering
\includegraphics[scale=0.5]{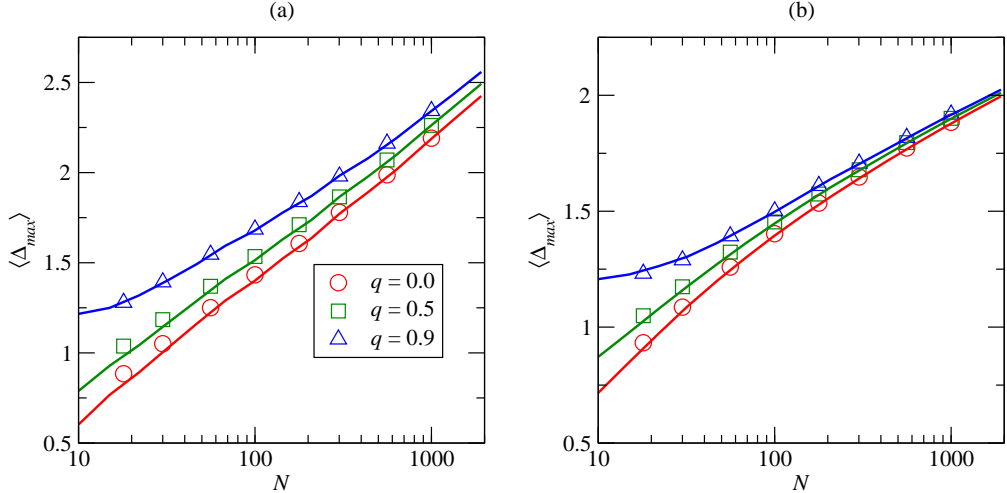}
\caption{(Color online)
Scaling of the expected maximum fluctuations $\langle\Delta_{\max}\rangle$ with system size $N$
for ensembles of (a) ER and (b) BA networks ($1000$ realizations). Open symbols correspond to simulating the
stochastic delay-differential equations Eq.~(\ref{nodeDiffEq}) with $\Delta t$$=$$0.001$)
Solid curves show predictions based on exact numerical diagonalization and employing Eqs.~(\ref{sigma_lambda}),
(\ref{sigma_i}), and (\ref{extremeDistributionDensity}).}
\label{scalingN}
\end{figure}

We found similar behavior for other prototypical networks [SW, ER
and BA], namely for $\tau$$<$$\tau_c$ the scaled distributions of
the extreme fluctuations converge to the FTG density
[Fig.~\ref{extremeDistributions}].
%%%%%%%%%%%%%%%%%%%%%%%%%%%%%%%%%%%%%%%%%%%%%%%%%%%%%%%%%%%%%%%%%%%%
%%%%%%%%%%%%%%%%%%%%%%%%%%%%%%%%%%%%%%%%%%%%%%%%%%%%%%%%%%%%%%%%%%%%
Also, our results for ER and BA networks indicate that the extreme
fluctuations $\langle\Delta_{\max}\rangle$ asymptotically approach a
logarithmic scaling with the system size $N$ [Fig.~\ref{scalingN}],
consistent with being governed by the FTG density.
%%%%%%%%%%%%%%%%%%%%%%%%%%%%%%%%
%%%%%%%%%%%%%%%%%%%%%%%%%%%%%%%%

\begin{figure}[t]
\vspace{1.0truecm}
\centering
\includegraphics[scale=0.6]{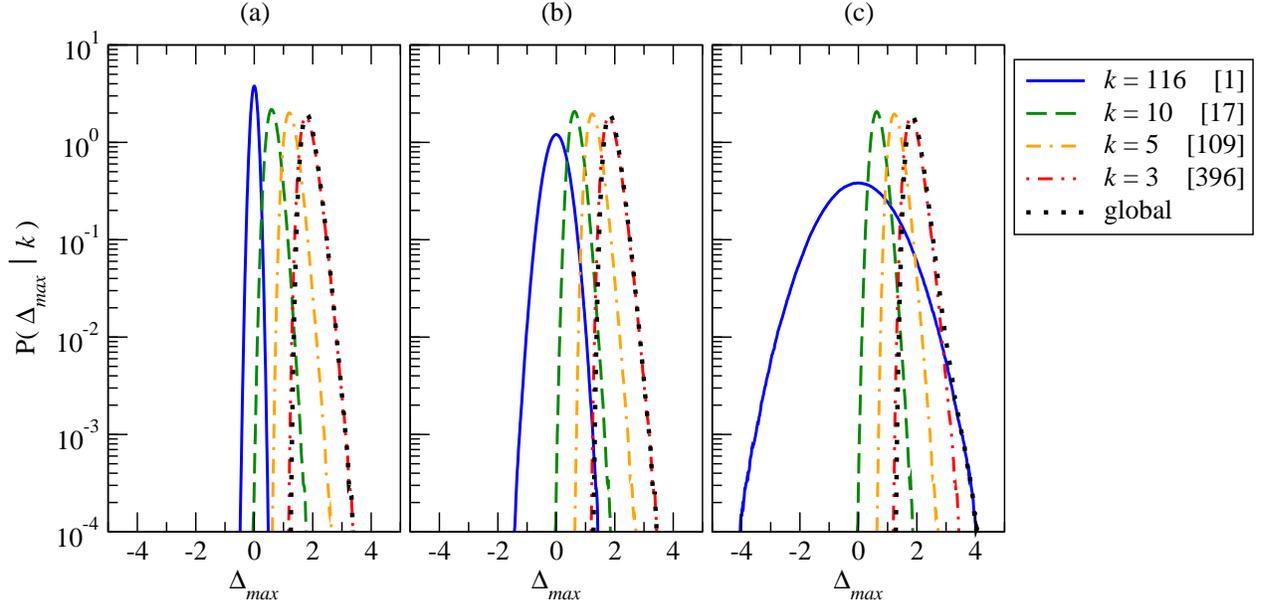}
\caption{(Color online)
Contribution of a few selected degree classes for a BA network (including $k_{\rm min}$$=$$3$ and $k_{\rm max}$$=$$116$)
with $N=1000$ and $\langle k\rangle \approx 6$ for (a) $q=0.0$, (b) $q=0.9$ and (c) $q=0.99$.
Results are obtained from numerical integration of the stochastic delay-differential equations Eq.~(\ref{nodeDiffEq})
with $\Delta t = 5 \times 10^{-6}$.
The numbers in brackets in the legends indicate the number of nodes in the corresponding degree class.}
\label{degreeClassDistributions}
\end{figure}

To provide further insights to the source of the extreme
fluctuations in BA networks, we also analyzed the distribution of
the extremes within each degree class. We have already seen that the
average size of fluctuations are the largest for small degrees,
except for near-critical delays
[Fig.~\ref{individualNodeVariances}]. When the delay approaches the
critical value for a given graph, the average size of the
fluctuations becomes a non-monotonic function of the degree (in a
mean-field sense, this behavior is naturally related to U-shape
scaling behavior of the fluctuations with the effective coupling
\cite{HuntPRL,HuntPRE}). In fact, there is a regime where the
fluctuations at the largest degree node are finite and are the
largest in the network. In parallel with the above observation,
sufficiently below the critical delay of a given graph, the extreme
fluctuations will almost always originate in the class of nodes with
the smallest degree: not only the average size of the fluctuations
is the largest here, but also this degree class has the largest
population (as given by the degree distribution). In this regime, it
is thus expected that the global distribution of the extreme
fluctuations will essentially overlap with the extremes within the
class of the minimum degree. Our simulations confirm this in
Fig.~\ref{degreeClassDistributions}(a) and (b). Further, as the
delay approaches its critical value in the given graph, the (single)
node with the largest degree will often give rise to the largest
fluctuations in the network. This is demonstrated in
Fig.~\ref{degreeClassDistributions}(c), showing that the tail of the
global extremes coincides with the (Gaussian) fluctuations at a
(single) node with the largest degree.

%\section{Mean-Field Approximation}

When the delay in a given network is not too close to the critical
delay, one can assume that the fluctuations at the nodes decouple.
This assumption works fairly well for complex networks with no
delays \cite{Guclu2004,Guclu_FNL,Guclu_Chaos2007}. (Note that
such assumption is ill-fated for low-dimensional spatial graphs
where a large correlation length governs the scaling.) Now we test
this hypothesis for complex networks with delays, and predict the
extreme limit distribution of the fluctuations.
The cumulative distribution of the fluctuations at a particular node
$i$ (with $x=\Delta_i$) can be expressed in terms of the error
function,
\begin{equation}
P^{<}_i(x) = \int_{-\infty}^{x} dx' p_{i}(x') =
\int_{-\infty}^{x} dx'\frac{e^{-x'^2/2\sigma_i^2}}{\sigma_i\sqrt{2\pi}} =
\frac{1}{2}\left\{ 1 + \frac{2}{\sqrt{\pi}}\int_{0}^{x/\sigma_i\sqrt{2}}dt e^{-t^2} \right\} =
\frac{1}{2}\left\{1 + \erf\left(\frac{x}{\sigma_i\sqrt{2}}\right)\right\} \;.
\label{individualNodeCumulative}
\end{equation}
Assuming independence of the nodes in determining the extremal
fluctuations, the cumulative distribution for the maximum fluctuation
during a given snapshot is
\begin{equation}
P^{{}<{}}_{\max}(x) \simeq \prod_{i = 1}^N P^{{}<{}}_i(x).
\label{maximumCumulative}
\end{equation}
The corresponding density from Eqs.~(\ref{individualNodeCumulative})
and (\ref{maximumCumulative}) is then
\begin{equation}
p_{\max}^{<}(x) = \frac{d}{dx}P^{<}_{\max}(x) \simeq
\sum_{i}^{N} p_{i}(x) \prod_{j\neq i} P^{{}<{}}_j(x) =
\frac{2^{1/2 - N}}{\sqrt{\pi}}\sum_{i=1}^{N} \left\{\frac{e^{-x^2/2\sigma_i^2}}{\sigma_i}\prod_{j \ne i}\left[1 + \erf\left(\frac{x}{\sigma_j\sqrt{2}}\right)\right]\right\}.
\label{extremeDistributionDensity}
\end{equation}
%%%%%%%%%%%%%%%%%%%%%%%%%%%%%%%%%%%%%%%%%%%%%%%%%%%%%%%%%%%%%%%%%%%%%%%%%%%%%%%%%%
The results based on the above approximation (together with those given by
the actual numerical simulations) are shown in Fig.~\ref{extremeDistributions} for the distribution,
and in Fig.~\ref{scalingN} for the average of the extremes.
%%%%%%%%%%%%%%%%%%%%%%%%%%%%%%%%%%%%%%%%%%%%%%%%%%%%%%%%%%%%%%%%%%%%%%%%%%%%%%%%%
Note that the validity of this approximation can break down when the
delay is sufficiently close to the critical delay so that
fluctuation at the few highest degree nodes can completely dominate
the extremes.
%%%%%%%%%%%%%%%%%%%%%%%%%%%%%%%%
%%%%%%%%%%%%%%%%%%%%%%%%%%%%%%%%
%%%%%%%%%%%%%%%%%%%%%%%%%%%%%%%%
%\begin{figure}[t]
%\vspace{1.0truecm}
%\centering
%\includegraphics[scale=0.4]{temp/ce_delayFrac_r_10_0.0001_compareZoom.eps}
%\caption{
%Rescaled theoretical (Eq (\ref{extremeDistributionDensity})) and simulated ($\Delta t = ????$) distributions for several delays near critical delay that show separate features from the low and high degree nodes.
%Also show distributions before rescaling!!!!
% }
%\label{criticalExtremeDistributions}
%\end{figure}
%%%%%%%%%%%%%%%%%%%%%%%%%%%%%%%%

Also, note that the fluctuations of the mode associated with the
largest eigenvalue assumes large oscillatory components for
$\lambda_{\max}\tau > 1/e$ \cite{HuntPRL,HuntPRE}. This manifests
with the greatest amplitude at the largest degree node with strong
oscillatory components. Once the delay is sufficiently close to the
critical delay so that these large-amplitude oscillations dominate
the extremes, an additional feature emerges in the distribution of
$\Delta_{\max}$. Here, there is a broader non-universal tail, which
originates from the finite but very large fluctuations at the node
with the largest degree [Fig.~\ref{degreeClassDistributions}(c)]. The
suppression of the contribution from low degree nodes is compounded
if more than one node has the network's maximum degree. Periodically,
when the oscillatory behavior of this node brings
it back near the mean $\bar h(t)$, the global extremes can still be
dominated by the FTG-distributed extremes among the lowest-degree
nodes.
%This contribution is observed as a peak in smaller fluctuations that
%shrinks in size as the delay is increased, while it slowly shifts to
%the right in the same linear trend as for smaller delays.

\section{The Width and the Extremes on Regular Lattices}

\begin{figure}[t]
\vspace{1.5truecm}
\centering
\includegraphics[scale=0.7]{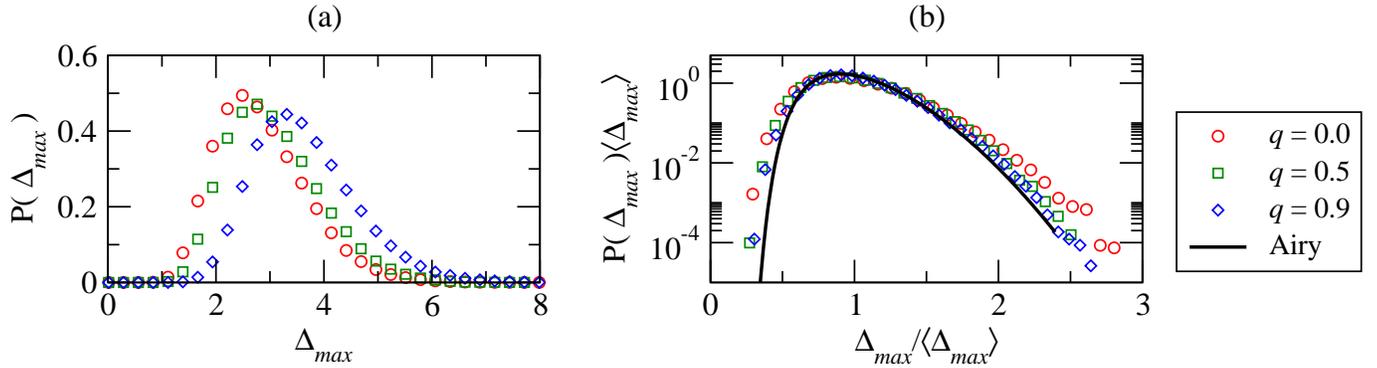}
\caption{(Color online)
(a) Extreme fluctuation distributions and (b) scaled
version of the same data for several delays for a regular
one-dimensional lattice with $N$$=$$1000$. Symbols are the results
found by numerically integrating Eq.~(\ref{nodeDiffEq}) using
$\Delta t = 5 \times 10^{-6}$. The solid line corresponds to the predicted asymptotic Airy limit distribution
\cite{Majumdar_2004,Majumdar_2005}.}
\label{regularDistributions}
\end{figure}

\begin{figure}[t]
\vspace{1.0truecm}
\centering
\includegraphics[scale=0.9]{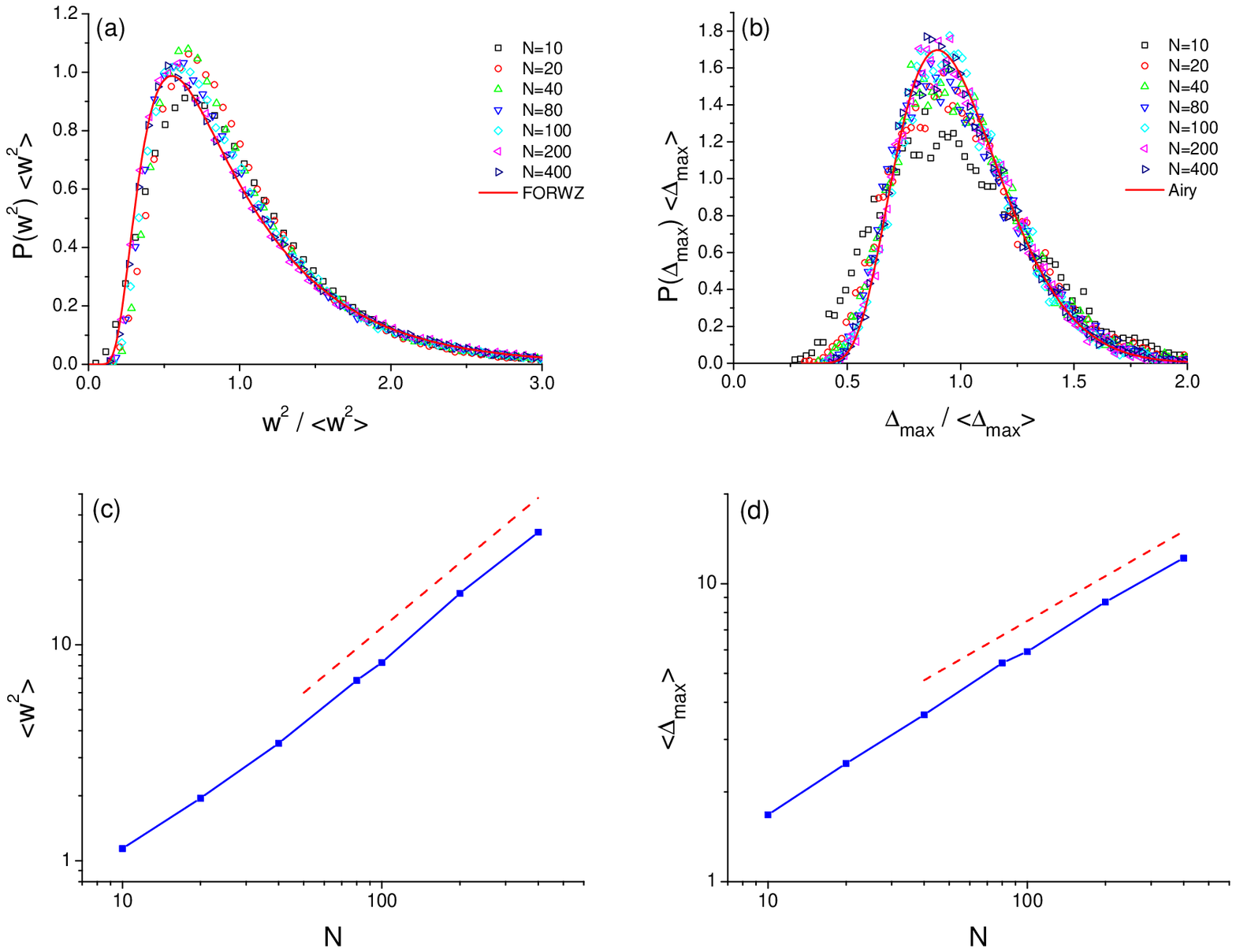}
\vspace{-0.50truecm}
\caption{(Color online)
(a) Finite-size behavior of the scaled width distribution for one-dimensional lattices for fixed time delay
$\tau=0.25$. The solid line corresponds to the predicted asymptotic
FORWZ limit distribution \cite{FORWZ_PRE1994}. (b) Finite-size
behavior of the scaled extreme fluctuation distribution for regular
one-dimensional lattices for the same system sizes and delay as in
(a). The solid line corresponds to the predicted asymptotic Airy limit
distribution \cite{Majumdar_2004,Majumdar_2005}.
(c) Finite-size behavior of the average width for the same system sizes and delay as in (a).
The dashed line is to guide the eyes, corresponding to the asymptotic theoretical prediction, $\langle w^2\rangle\sim N$.
(d) Finite-size behavior of the extreme fluctuations for the same system sizes and delay as in (b).
The dashed line is to guide the eyes, corresponding to the asymptotic theoretical prediction, $\langle \Delta_{\max}\rangle\sim N^{1/2}$.}
\label{regular_fss}
\end{figure}

Finally, it is worthwhile to  contrast the steady-state scaling
behavior of the extremes and their distributions in complex networks to those on regular
lattices. For regular $d$-dimensional lattices the largest
eigenvalue of the Laplacian is {\it independent} of the system size.
Thus, for a fixed delay $\tau$, if the system is synchronizable for
a particular system size, it is synchronizable for all system sizes,
$\lambda_{\max}\tau<\pi/2$, and the usual $N\to\infty$ thermodynamic
limit can be considered with no delay-induced instability. Further,
as $N\to\infty$, the arbitrarily small eigenvalues of the Laplacian
($\lambda_{\min}\sim N^{-2/d}$) will dominate the sum in
Eq.~(\ref{exactWidth}), just like they do in systems with no delay.
Hence, one can expect that the scaling of the width, the  extremes
and their asymptotic limit distribution in the synchronizable regime
will be identical to those with no delay, governed by a diverging
correlations and long-wavelength modes (associated with the
arbitrarily small eigenvalues of the Laplacian). For illustration,
we considered one-dimensional lattices (with nearest-neighbor
coupling) with delays [Fig.~\ref{regularDistributions}]. For
completeness, we studied the detailed finite-size behavior of both
the steady-state width distribution $P(w^2)$ and the distribution of
the extremes $P(\Delta_{\max})$. The results of the numerical
integration of the systems with delays (but within the
synchronizable regime, $\lambda_{\max}\tau<\pi/2$) show that the
asymptotic limit distributions of the width and the extreme
fluctuations approach those of the delay-free systems, the FORWZ
distribution \cite{FORWZ_PRE1994} and the Airy distribution
\cite{Majumdar_2004,Majumdar_2005}, respectively
[Fig.~\ref{regular_fss}].
%%%%%%%%%%%%%%%%%%%%%%%%%%%%%%%%
%%%%%%%%%%%%%%%%%%%%%%%%%%%%%%%%

%%%%%%%%%%%%%%%%%%%%%%%%%%%%%%%%
%%%%%%%%%%%%%%%%%%%%%%%%%%%%%%%%

\section{Exploring the Effects of Nonlinear Couplings}

While in our current work the focus has been to understand the effects of time delays on stochastic systems with liner couplings, we performed some explorations on the nonlinear effects. We considered the stochastic equation
%%%%%%%%%%%%%%%%%%%%%%%%%%%%%%%%%%%%%%%%%%%%%%%%%%%%%%%%%%%%%%%%%%%%%%%%%%%%%%%%%%%%%%%%%%%%%%%%%%%%%%
\begin{equation}
\partial_t h_i(t) = -\sum_{j = 1}^N C_{ij}(h_i - h_j) + \nu\sum_{j,k = 1 (j<k)}^N C_{ij}C_{ik}(h_j - h_i)(h_k - h_i) + \eta_i(t) \;,
\label{network_KPZ}
\end{equation}
%%%%%%%%%%%%%%%%%%%%%%%%%%%%%%%%%%%%%%%%%%%%%%%%%%%%%%%%%%%%%%%%%%%%%%%%%%%%%%%%%%%%%%%%%%%%%%%%%%%%%%
where the effects of time delays are captured, as before, by replacing $\{h_{l}(t)\}_{l=1}^{N}$ by $\{h_{l}(t-\tau)\}_{l=1}^{N}$ in the right-hand side of the above stochastic differential equation. Such nonlinear terms can be motivated by, for example, coarse-graining local growth processes in networks \cite{LaRocca_PRE2008,Korniss_PRL2000,Guclu_PRE2006,Guclu_Chaos2007,Guclu_unpublished2008} (e.g., where only local network-neighborhood minima can progress). Note that in one dimension, the above stochastic equation reduces to
%%%%%%%%%%%%%%%%%%%%%%%%%%%%%%%%%%%%%%%%%%%%%%%%%%%%%%%%%%%%%%%%%%%%%%%%%%%%%%%%%%%%%%%%%%%%%%%%%%%%%%%%
\begin{equation}
\partial_t h_i(t) = -(2h_i - h_{i+1} -h_{i-1}) + \nu(h_{i+1} - h_i)(h_{i-1} - h_i) + \eta_i(t) \;.
\label{1d_KPZ}
\end{equation}
%%%%%%%%%%%%%%%%%%%%%%%%%%%%%%%%%%%%%%%%%%%%%%%%%%%%%%%%%%%%%%%%%%%%%%%%%%%%%%%%%%%%%%%%%%%%%%%%%%%%%%%%
The above equation is just a (somewhat unconventional) discretization of the well-known KPZ equation \cite{KPZ,Barabasi},
%%%%%%%%%%%%%%%%%%%%%%%%%%%%%%%%%%%%%%%%%%%%%%%%%%%%%%%%%%%%%%%%%%%%%%%%%%%%%%%%%%%%%%%%%%%%%%%%%%%%%%%%
\begin{equation}
\partial_t h(x,t) = \nabla^2 h -\nu (\nabla h)^2 + \eta_i(t) \;,
\label{1d_KPZ_continuum}
\end{equation}
%%%%%%%%%%%%%%%%%%%%%%%%%%%%%%%%%%%%%%%%%%%%%%%%%%%%%%%%%%%%%%%%%%%%%%%%%%%%%%%%%%%%%%%%%%%%%%%%%%%%%%%%
which can be easily seen by taking the naive continuum limit in Eq.~(\ref{1d_KPZ}), $h_{i\pm 1}(t)=h(x,t)\pm\partial h/\partial x+\ldots$, and keeping only the leading-order derivatives.

%%%%%%%%%%%%%%%%%%%%%%%%%%%%%%%%
\begin{figure}[t]
\vspace{1.0truecm}
\centering
\includegraphics[scale=0.55]{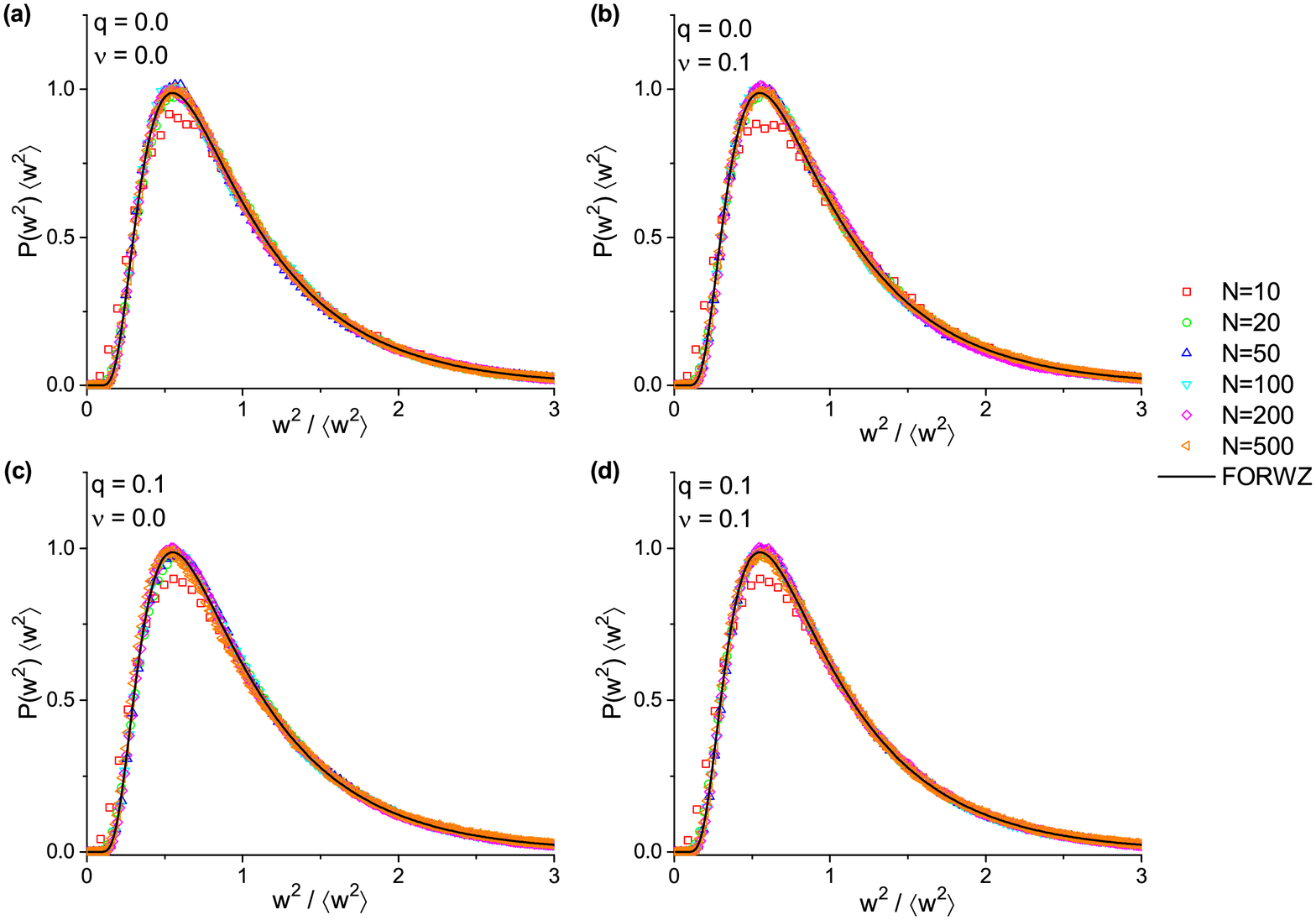}
%{Pw2_ring.eps}
\vspace{-0.00truecm}
\caption{(Color online)
Finite-size behavior of the scaled width distribution for one-dimensional lattices with nonlinear coupling $\nu$ [Eq.~(\ref{1d_KPZ})] and relative time delay $q=\tau/\tau_{\rm c}$.
(a) $\nu=0.0$, $q=0.0$;
(b) $\nu=0.1$, $q=0.0$;
(c) $\nu=0.0$, $q=0.1$;
(d) $\nu=0.1$, $q=0.1$.
The solid line corresponds to the predicted asymptotic FORWZ limit distribution \cite{FORWZ_PRE1994}.}
\label{Pw2_1d_nonlin_fss}
\end{figure}
%%%%%%%%%%%%%%%%%%%%%%%%%%%%%%%%

%%%%%%%%%%%%%%%%%%%%%%%%%%%%%%%%
\begin{figure}[t]
\vspace{1.0truecm}
\centering
\includegraphics[scale=0.55]{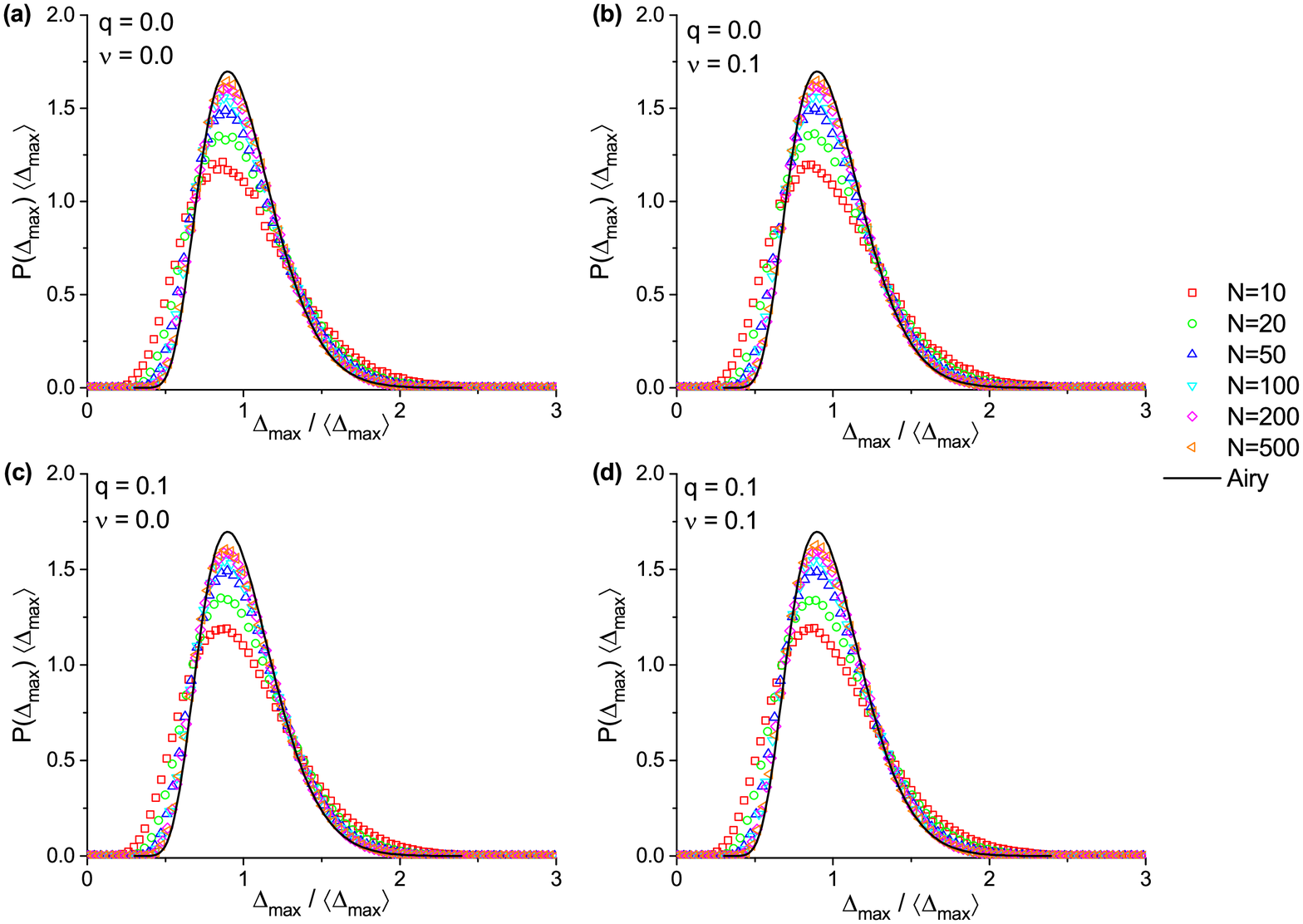}
%{PDmax_Ring.eps}
\vspace{-0.00truecm}
\caption{(Color online)
Finite-size behavior of the scaled distribution of the extreme fluctuation above the mean for regular one-dimensional lattices with nonlinear coupling $\nu$ [Eq.~(\ref{1d_KPZ})] and relative time delay $q=\tau/\tau_{\rm c}$.
(a) $\nu=0.0$, $q=0.0$;
(b) $\nu=0.1$, $q=0.0$;
(c) $\nu=0.0$, $q=0.1$;
(d) $\nu=0.1$, $q=0.1$.
The solid line corresponds to the predicted asymptotic Airy limit distribution \cite{Majumdar_2004,Majumdar_2005}.}
\label{PDmax_1d_nonlin_fss}
\end{figure}
%%%%%%%%%%%%%%%%%%%%%%%%%%%%%%%%

%%%%%%%%%%%%%%%%%%%%%%%%%%%%%%%%
\begin{figure}[t]
\vspace{1.0truecm}
\centering
\includegraphics[scale=0.55]{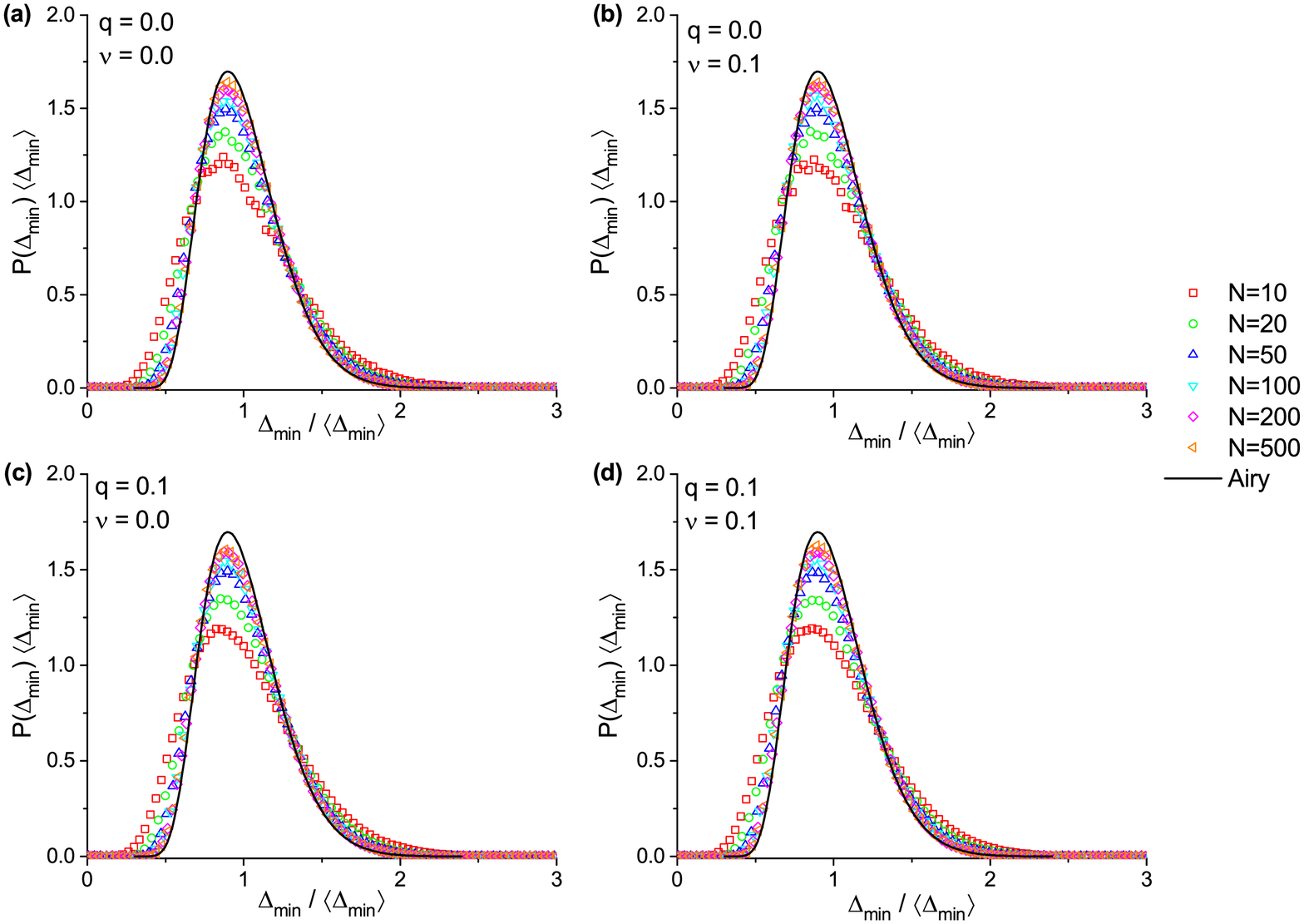}
%{PDmin_Ring.eps}
\vspace{-0.00truecm}
\caption{(Color online)
Finite-size behavior of the scaled distribution of the extreme fluctuation below the mean for regular one-dimensional lattices with nonlinear coupling $\nu$ [Eq.~(\ref{1d_KPZ})] and relative time delay $q=\tau/\tau_{\rm c}$.
(a) $\nu=0.0$, $q=0.0$;
(b) $\nu=0.1$, $q=0.0$;
(c) $\nu=0.0$, $q=0.1$;
(d) $\nu=0.1$, $q=0.1$.
The solid line corresponds to the predicted asymptotic Airy limit distribution \cite{Majumdar_2004,Majumdar_2005}.}
\label{PDmin_1d_nonlin_fss}
\end{figure}
%%%%%%%%%%%%%%%%%%%%%%%%%%%%%%%%

First, we studied one-dimensional regular lattices (with nearest-neighbor connections and periodic boundary conditions). In what follows, we parameterized the delay relative to the critical delay of the {\em linear} system for reference, $q\equiv\tau/\tau_{c}$. Numerically integrating the time-discretized version of the stochastic differential equation Eq.~(\ref{1d_KPZ}) (see Supplemental Material for more details), we have found that for sufficiently small values of the nonlinear coupling $\nu$ and time delays, the system reaches a steady state with a finite width for finite systems. The width distribution [Fig.~\ref{Pw2_1d_nonlin_fss}] and the distribution of the extremes for both above [$\Delta_{\max}=h_{\max}-\bar{h}$, Fig.~\ref{PDmax_1d_nonlin_fss}] and below the mean [$\Delta_{\min}=\bar{h}-h_{\min}$, Fig.~\ref{PDmin_1d_nonlin_fss}] approach the FORWZ and the Airy distribution, respectively, similar to the case of pure linear couplings.
%%%%%%%%%%%%%%%%%%%%%%%%%%%%%%%%
\begin{figure}[t]
\vspace{1.0truecm}
\centering
\includegraphics[scale=0.55]{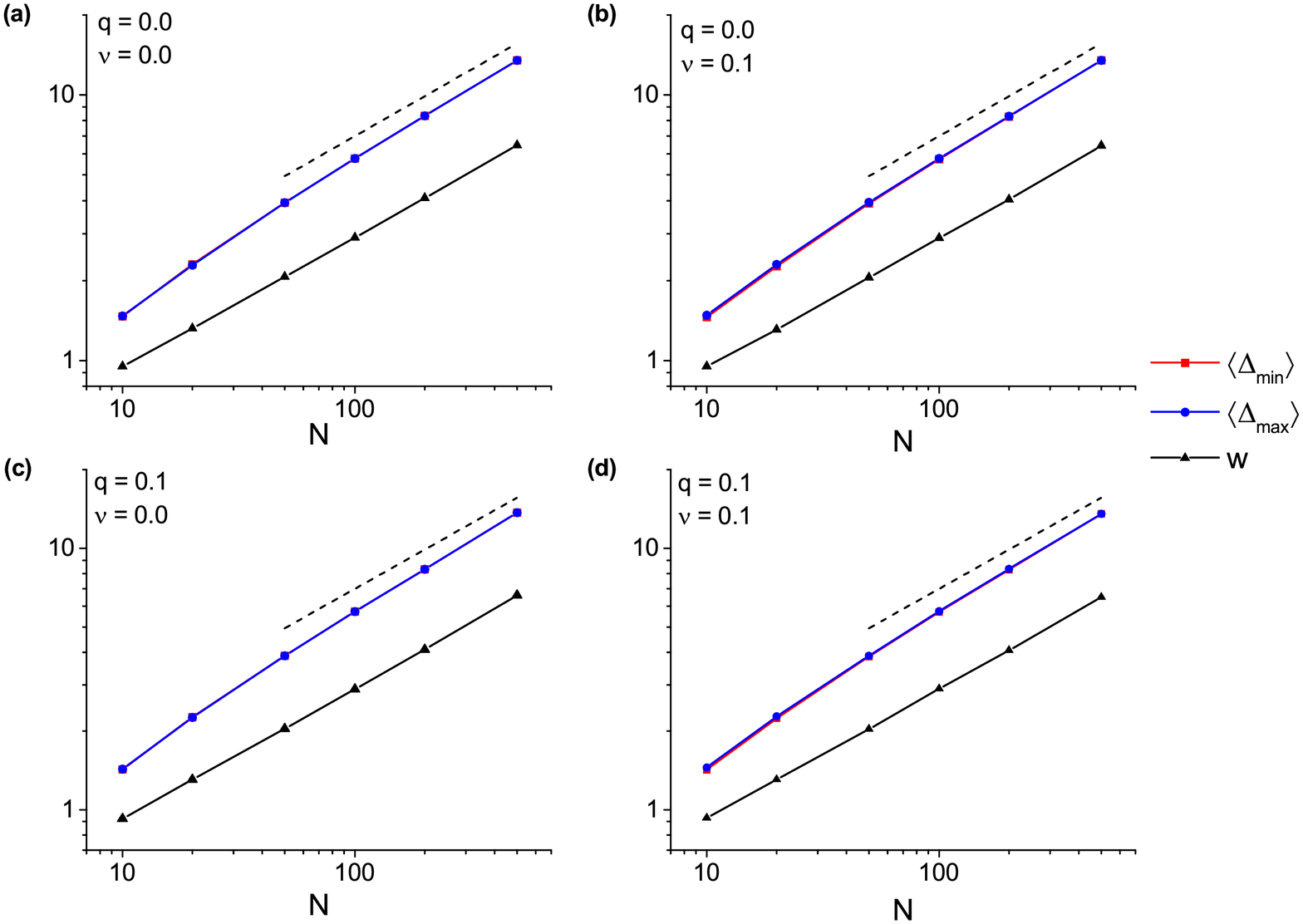}
%{N_Ring_log.eps}
\vspace{-0.00truecm}
\caption{(Color online)
Scaling of the average width $w\equiv\sqrt{\langle w^2\rangle}$ and the expected extreme fluctuations $\Delta_{\max}$ and $\Delta_{\min}$ (the expected largest fluctuations above and below the mean, respectively) with the system size in regular one-dimensional lattices with nonlinear coupling $\nu$ [Eq.~(\ref{1d_KPZ})] and relative time delay $q=\tau/\tau_{\rm c}$.
The dashed lines are to guide the eyes, corresponding to the scaling $\sim N^{1/2}$. Note the log-log scales.
(a) $\nu=0.0$, $q=0.0$;
(b) $\nu=0.1$, $q=0.0$;
(c) $\nu=0.0$, $q=0.1$;
(d) $\nu=0.1$, $q=0.1$.}
\label{Avg_1d_nonlin_fss}
\end{figure}
%%%%%%%%%%%%%%%%%%%%%%%%%%%%%%%%
Further, the average width and the extremes (both above and below the mean height) scale as $N^{1/2}$ with the system size [Fig.~\ref{Avg_1d_nonlin_fss}].

Investigating the behavior of the width for larger values of the nonlinear coupling, it is clear the synchronization profile {\em can} diverge (the system becomes unstable), even for zero time delay (see Supplemental Material). We checked and tested that this instability is {\em not} an artifact of the finite time difference $\Delta t$, but rather it is the result of the non-linear term on discrete lattices in Eq.~(\ref{1d_KPZ}).
Indeed, it has been well documented \cite{Majumdar_2005,DasSarma_PRE1996,DasSarma_PRE1997,Newman_JPA1996,Shin_PRE1998a,Shin_PRE1998b} that even conventional lattice discretization schemes of the KPZ nonlinearity can give rise to instability in a noisy environment (even though its spatial continuum limit is stable and the nonlinear term, in fact, exactly cancels by symmetry in the stationary state \cite{KPZ,Barabasi,EW}). Our observed behavior, induced by the nonlinear KPZ term in Eq.~(\ref{1d_KPZ}), is just another example for such instability, intrinsic on discrete structures, such as lattices.

%This is interesting, as the nonlinear effects in the {\em one-dimensional} KPZ equation (at least in the exact continuum limit) are exactly eliminated
%in the steady state by an inherent symmetry of the system \cite{KPZ,Barabasi}; the steady-state behavior, in turn, is equivalent to that described by
%the linear Edwards-Wilkinson equation \cite{EW,Barabasi}. Thus, the non-linear term in Eq.~(\ref{1d_KPZ}), even though being a particular
%discretization of the KPZ equation, does not guarantee stability for an arbitrary coupling strength.

\begin{figure}[t]
\vspace{1.0truecm}
\centering
\includegraphics[scale=0.55]{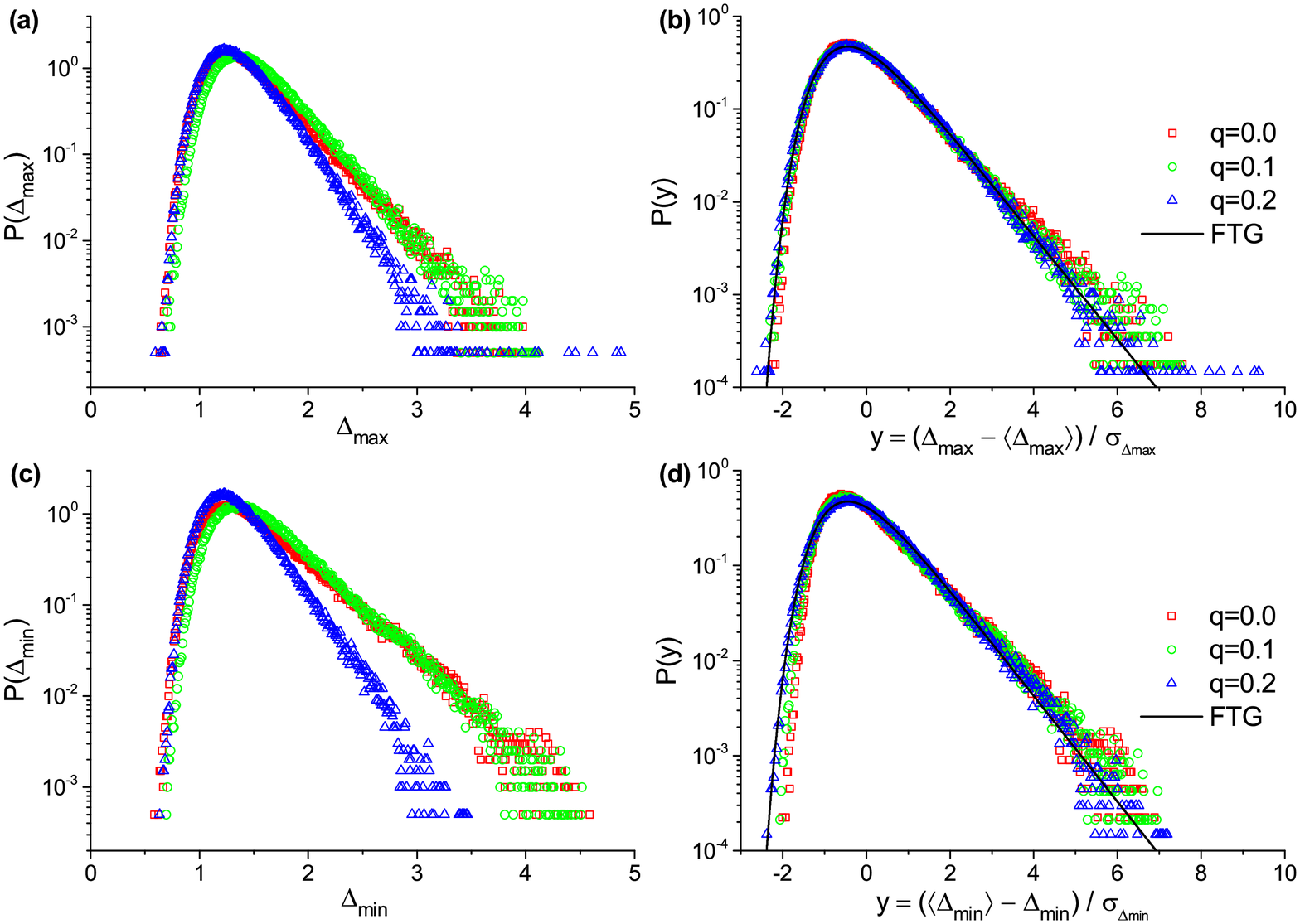}
%{PDmaxmin.eps}
\vspace{-0.00truecm}
\caption{(Color online)
Extreme fluctuation distribution (in the quasi-stationary period) in ER networks with $N=100$, $\langle k\rangle\approx 6$, nonlinear coupling $\nu=0.1$ [Eq.~(\ref{network_KPZ})], and various relative time delays $q=\tau/\tau_{\rm c}$.
(a) Distributions of the extreme fluctuations above the mean, $\Delta_{\max}=h_{\max}-\bar{h}$;
(b) Scaled distributions of the extreme fluctuations above the mean (with zero mean and unit variance);
(c) Distributions of the extreme fluctuations below the mean, $\Delta_{\min}=\bar{h}-h_{\min}$;
(d) Scaled distributions of the extreme fluctuations below the mean (with zero mean and unit variance).
The solid line in (b) and (d) corresponds to the FTG limit distribution [Eq.~(\ref{FTG_density})].}
\label{PDmaxmin_ER_nonlin}
\end{figure}

In networks, we found similar behavior. For example, in ER networks, we found that the nonlinear term in Eq.~(\ref{network_KPZ}) gives rise to a diverging width even for our smallest value of the nonlinear coupling strength $\nu$, even in the absence of time delays (see Supplemental Material). We again checked and confirmed that the lack of stability in the presence of nonlinear couplings is {\em not} the result of insufficient time discretization of the stochastic differential equation (\ref{network_KPZ}) (see Supplemental Material). For small values of the nonlinear coupling $\nu$, there exist, however, a long ``quasi-stationary" period before fluctuations begin to diverge, where we analyzed the statistical properties of the extremes. We have found that during this quasi-stationary period, the scaled distributions of the extremes are reasonably well-described by the FTG limit densities [Fig.~\ref{PDmaxmin_ER_nonlin}].
%%%%%%%%%%%%%%%%%%%%%%%%%%%%%%%
%%%%%%%%%%%%%%%%%%%%%%%%%%%%%%%%

\begin{figure}[t]
\vspace{1.0truecm}
\centering
\includegraphics[scale=0.55]{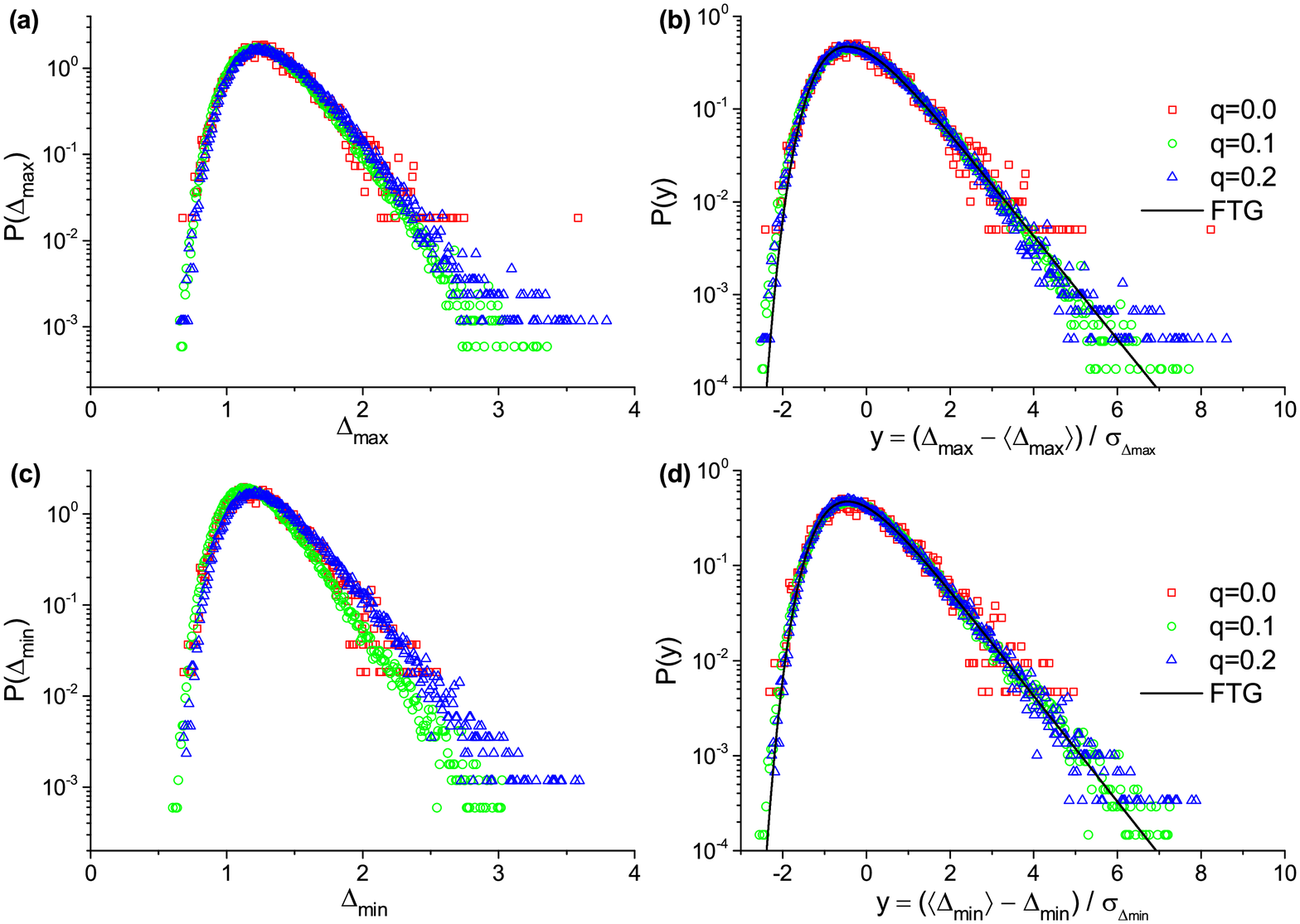}
%{PDmaxmin_SW.eps}
\vspace{-0.00truecm}
\caption{(Color online)
Extreme fluctuation distribution (in the quasi-stationary period) in SW networks with $N=100$, $\langle k\rangle\approx 6$, nonlinear coupling $\nu=0.1$ [Eq.~(\ref{network_KPZ})], and various relative time delays $q=\tau/\tau_{\rm c}$.
(a) Distributions of the extreme fluctuations above the mean, $\Delta_{\max}=h_{\max}-\bar{h}$;
(b) Scaled distributions of the extreme fluctuations above the mean (with zero mean and unit variance);
(c) Distributions of the extreme fluctuations below the mean, $\Delta_{\min}=\bar{h}-h_{\min}$;
(d) Scaled distributions of the extreme fluctuations below the mean (with zero mean and unit variance).
The solid line in (b) and (d) corresponds to the FTG limit distribution [Eq.~(\ref{FTG_density})].}
\label{PDmaxmin_SW_nonlin}
\end{figure}

For SW networks, while the fluctuations (and the width) are smaller than those than on one-dimensional lattices (at least during the quasi-stationary period), the synchronization landscape eventually becomes unstable for sufficiently strong nonlinear coupling, with or without delays (see Supplemental Material). We also observe that the average width in the quasi-stationary period is decreasing with increasing average degree $\langle k\rangle$, but at the same time, the duration of the quasi-stationary period is decreasing with increasing $\langle k\rangle$ (see Supplementary Material). Nevertheless, in the stationary (or quasi-stationary) state, the scaled distributions of the extremes are well-described by the FTG limit densities [Fig.~\ref{PDmaxmin_SW_nonlin}].
%%%%%%%%%%%%%%%%%%%%%%%%%%%%%%%
%%%%%%%%%%%%%%%%%%%%%%%%%%%%%%%%

It is clear that we have only begun to scratch the surface of the complexity of the behavior as a result of nonlinear couplings in the stochastic delay differential equations in networks. Among the important questions that one shall investigate are the effects of the strength of nonlinear coupling, time delay, and network size on the length of the quasi-stationary period. It is also clear from our explorations that uncontrolled expansions of local growth processes (i.e., naive coarse graining) may result in nonlinear terms in the resulting stochastic differential equations (with or without delays) which give rise to instability and diverging width in the synchronization landscape. This instability is ``real" (i.e., not an artifact of time discretization) as far as the numerical integration of the stochastic differential equation is concerned, but may not be present in the actual physical systems with the original microscopic rules \cite{LaRocca_PRE2008,Korniss_PRL2000,Guclu_PRE2006,Guclu_Chaos2007}.

\section{Summary}

We have demonstrated that the extreme fluctuations in stochastic
coordination or synchronization problems with time delays (with
short-tailed node-level noise and within the linearized
approximation) can fall in two main classes. In complex or random
networks (e.g., ER, SF, or SW graphs), if the system is sufficiently
large but synchronizable ($\lambda_{\max}\tau<\pi/2$), the
distribution of the extremes is governed by the FTG distribution,
while the average size of the largest fluctuations does not grow
faster than logarithmic. This type of scaling behavior can be
understood as the fluctuations at the nodes are only weakly
correlated, hence traditional extreme-value limit theorems apply. In
contrast, in spatial graphs, fluctuations at the nodes are strongly
correlated. As demonstrated on one-dimensional regular rings, the
distribution of the extremes approaches the Airy limit distribution,
while the average size of the largest fluctuations will scale as the
width itself, e.g., as a power law in one dimension.

Finally, we have performed some explorations on the effects on nonlinear couplings in the stochastic delay differential equations in networks.
Our results indicate the generalized KPZ nonlinearity in discrete structures (networks) can ultimately give rise to instability. Even in that case, however, during a quasi-stationary period (before the width diverges), the statistics of the extreme fluctuations are well-described by the Airy and FTG densities, in one dimension and in random ER/SW graphs, respectively. Clearly, future investigations are needed to precisely characterize the stability conditions in the presence of nonlinear couplings in networks with (and without) time delays.

\section*{Acknowledgments}
This research was supported in part
by DTRA Award No. HDTRA1-09-1-0049,
by NSF Grant No. DMR-1246958, and by
the Army Research Laboratory under Cooperative Agreement Number W911NF-09-2-0053 (the ARL Network Science CTA).
The views and conclusions contained in this document are those of the authors and should not be interpreted as
representing the official policies, either expressed or implied, of the Army Research Laboratory or the US Government.

\newpage

%\newcommand{\erf}{{\rm erf}}
%\renewcommand{\thefigure}{S\arabic{figure}}
%\renewcommand{\thetable}{S\arabic{table}}

%For Supplemental Material figure/equation/section numbering
\renewcommand{\thefigure}{S\arabic{figure}}\setcounter{figure}{0}
\renewcommand{\theequation}{S\arabic{equation}}\setcounter{equation}{0}
% (so that equations are numbered (S1), (S2), ...
\renewcommand{\thesection}{S\arabic{section}}\setcounter{section}{0}

\begin{center}
{\bf \large Supplemental Material}
\end{center}

\vspace*{2cm}

%D. Hunt\footnote{present address: Department of Biomathematics, David Geffen School of Medicine at UCLA, Los Angeles, CA 90095, USA}
%\affiliation{Department of Physics, Applied Physics, and Astronomy}
%\affiliation{Network Science and Technology Center}
%
%
%F. Moln\'ar, Jr.\footnote{present address: Department of Physics and Astronomy, Northwestern University, 2145 Sheridan Rd., Evanston, IL 60208, USA}
%\affiliation{Department of Physics, Applied Physics, and Astronomy}
%\affiliation{Network Science and Technology Center}
%
%B.K. Szymanski
%\affiliation{Department of Computer Science \\
%Rensselaer Polytechnic Institute, 110 8$^{th}$ Street, Troy, NY 12180--3590, USA}
%\affiliation{Network Science and Technology Center}
%
%\author{G. Korniss\footnote{Corresponding author. korniss@rpi.edu}
%\affiliation{Department of Physics, Applied Physics, and Astronomy}
%\affiliation{Network Science and Technology Center}
%

%%%%%%%%%%%%%%%%%%%%%%%%%%%%%%%%

%\date{\today}
%\maketitle

\begin{figure}[ht]
\centering
\includegraphics[width=\textwidth]{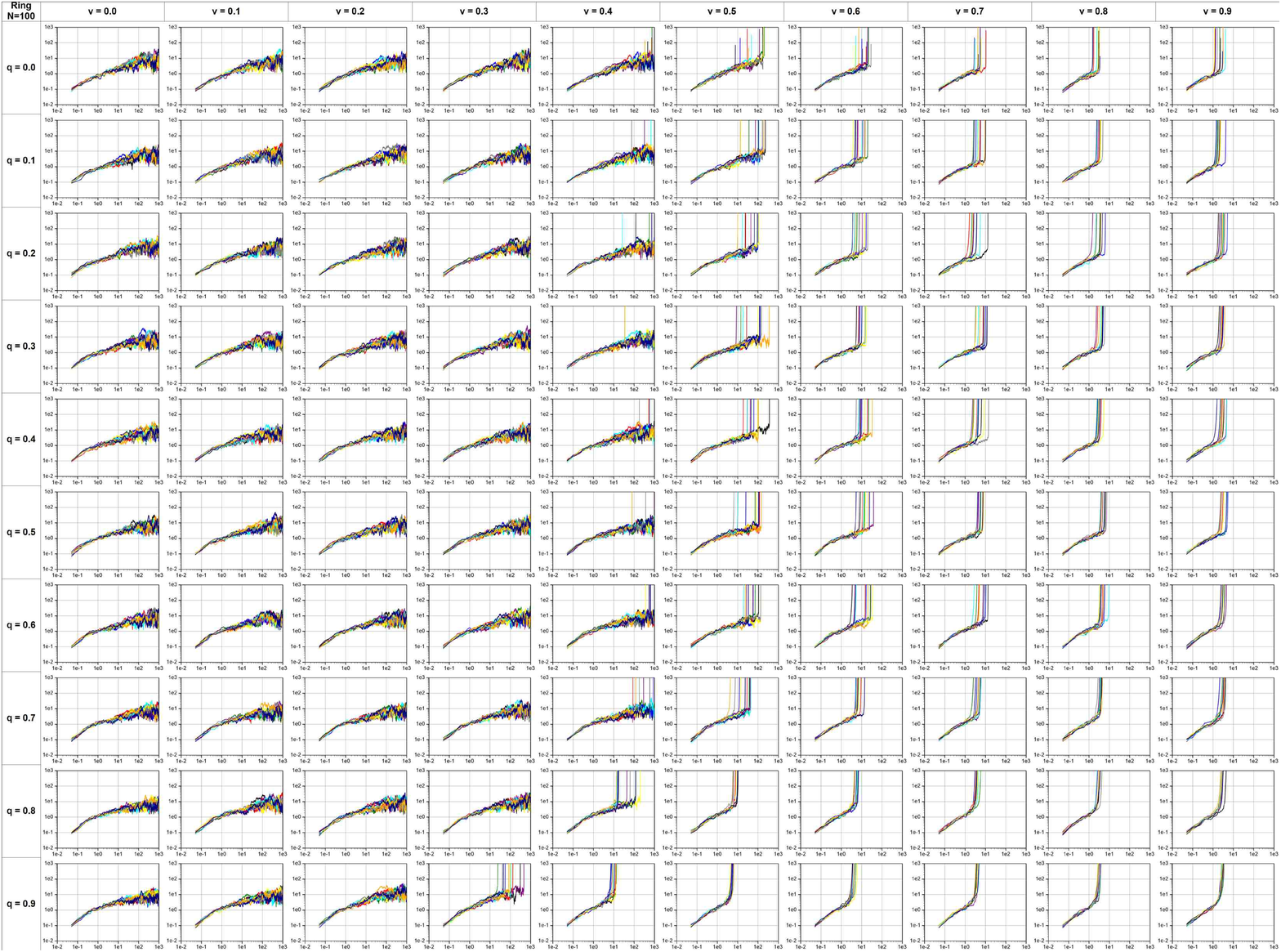}
\caption{Graphical table of the evolution of the width over time, in a one-dimensional regular lattice (with periodic boundary conditions) with $N$$=$$100$ for KPZ-like coupling [Eq.~(\ref{1d_KPZ_dt})]. Each tile shows the width $\langle w^2 \rangle$ as a function of time, using identical scales in each tile. Data was obtained by numerically integrating Eq.~(\ref{1d_KPZ_dt}) with $\Delta t=10^{-3}$.
Colors represent $10$ distinct realizations of noise. Each row corresponds to the indicated value of time delay $q = \tau / \tau_c$, and each column corresponds to the indicated nonlinear coupling strength $\nu$. A high-resolution version of this figure is provided separately in TIFF format among the Supplemental Materials (W2vsTimeMap\_Ring.tif).}
\label{fig-map_1d}
\end{figure}

\begin{figure}[ht]
\centering
\includegraphics[width=4in]{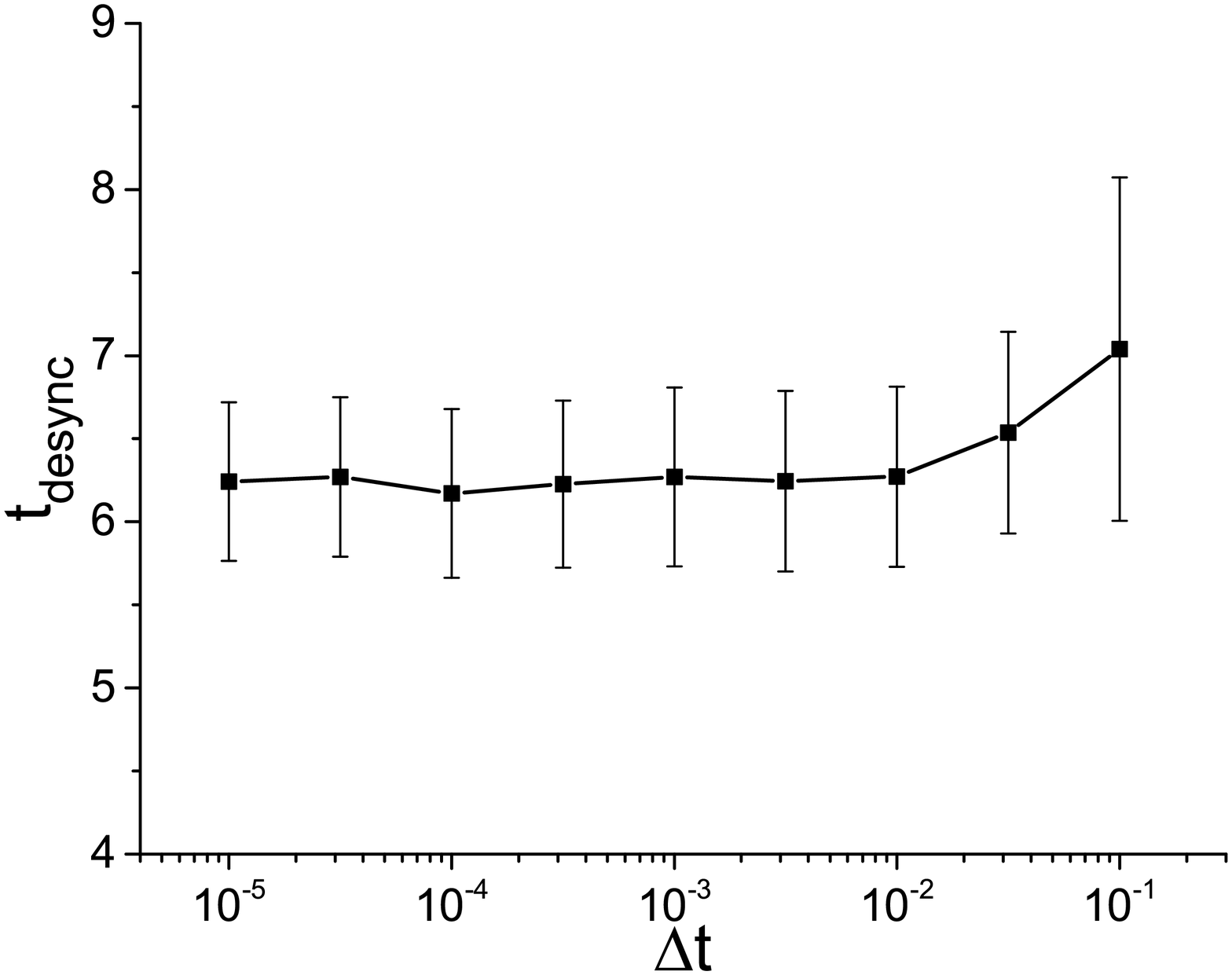}
\caption{Average time to desynchronization for KPZ-like coupling [Eq.~(\ref{1d_KPZ_dt})]
at various $\Delta t$ time steps of integration on a one-dimensional regular lattice (with periodic boundary conditions) with $N$$=$$100$, $\nu = 0.8$, $q = 0.8$. Error bars represent standard deviation, sampled over $100$ realizations.}
\label{fig-numint_1d}
\end{figure}

\begin{figure}[ht]
\centering
\includegraphics[width=4in]{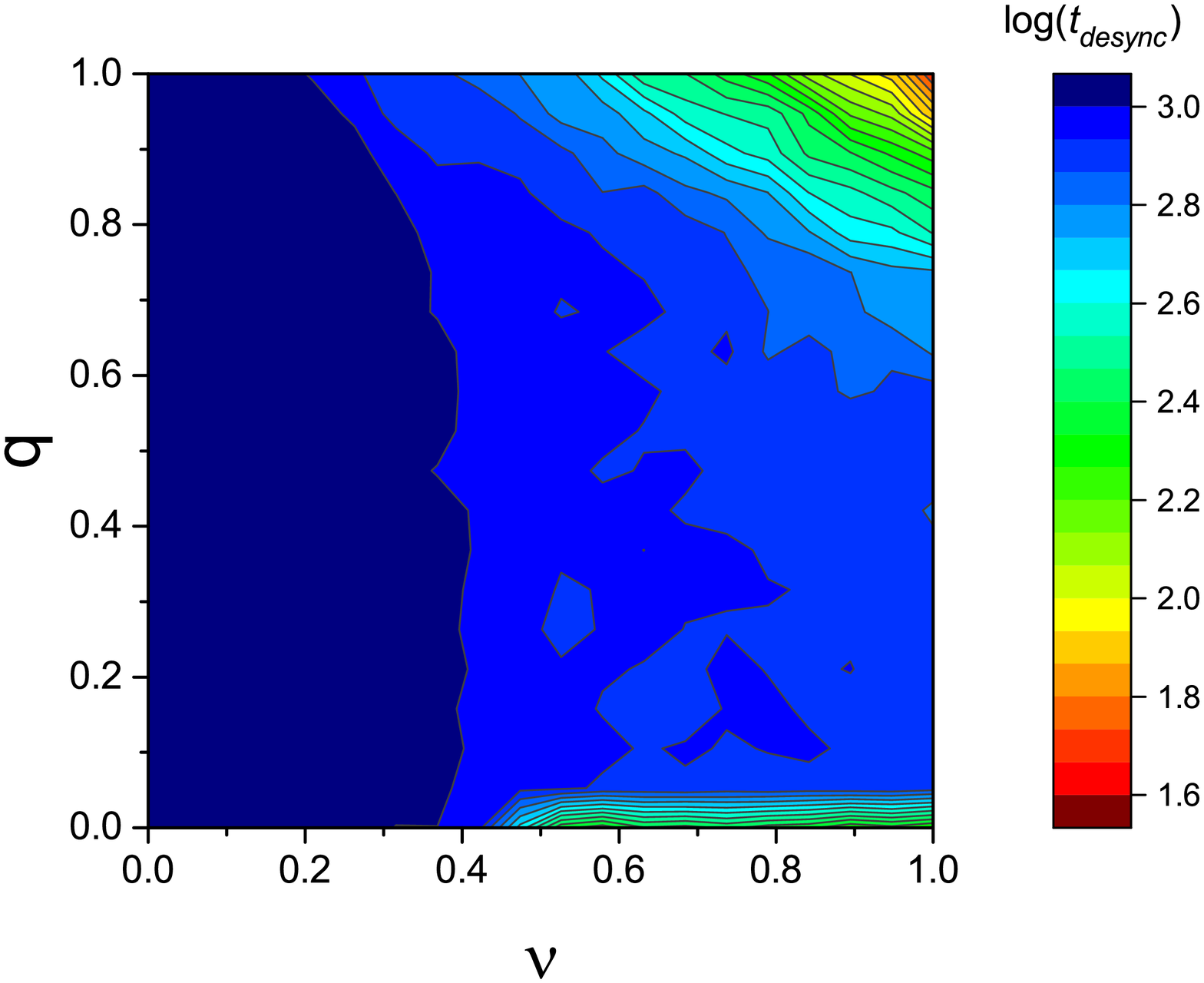}
\caption{Time to reach desynchronization for KPZ-like coupling [Eq.~(\ref{1d_KPZ_dt})], as a function  of the delay $q = \tau / \tau_c$ and nonlinear coupling strength $\nu$, on a one-dimensional regular lattice (with periodic boundary conditions) with $N$$=$$100$.
Data was obtained by numerically integrating Eq.~(\ref{1d_KPZ_dt}) with $\Delta t=10^{-3}$.}
\label{fig-desync_1d}
\end{figure}

\begin{figure}[ht]
\centering
\includegraphics[width=\textwidth]{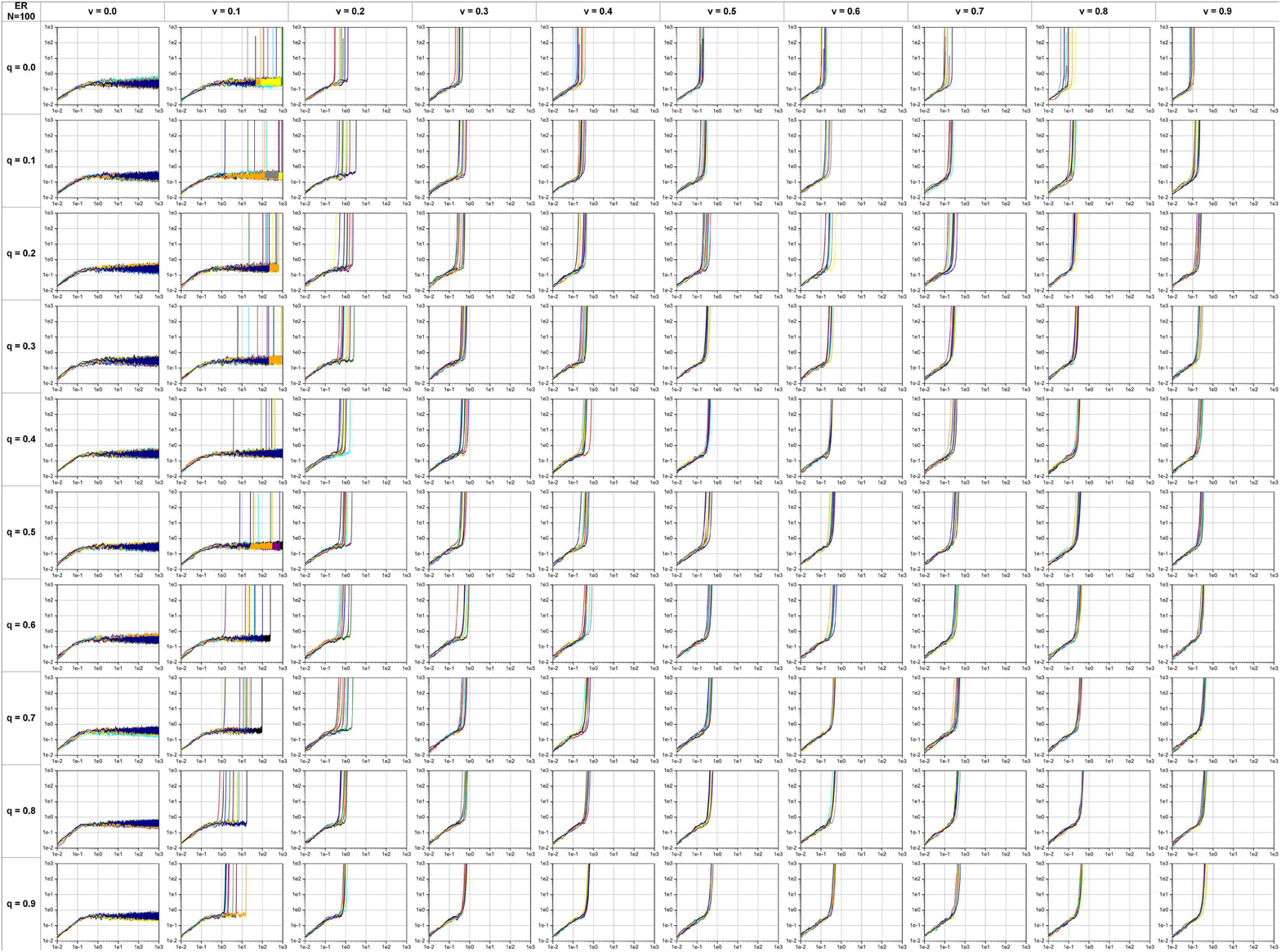}
\caption{ Graphical table of the evolution of the width over time, in ER networks with $N$$=$$100$ and $\langle k\rangle \approx 6$ using generalized KPZ coupling [Eq.~(\ref{network_KPZ_dt})]. Each tile shows the width $\langle w^2 \rangle$ as a function of time, using identical scales in each tile.
Data was obtained by numerically integrating Eq.~(\ref{network_KPZ_dt}) with $\Delta t=10^{-4}$.
Colors represent $10$ distinct realizations of noise. Each row corresponds to the indicated value of time delay $q = \tau / \tau_c$, and each column corresponds to the indicated nonlinear coupling strength $\nu$. A high-resolution version of this figure is provided separately in TIFF format among the Supplemental Materials (W2vsTimeMapER\_k6.tif).}
\label{fig-map_ER}
\end{figure}

\begin{figure}[ht]
\centering
\includegraphics[width=4in]{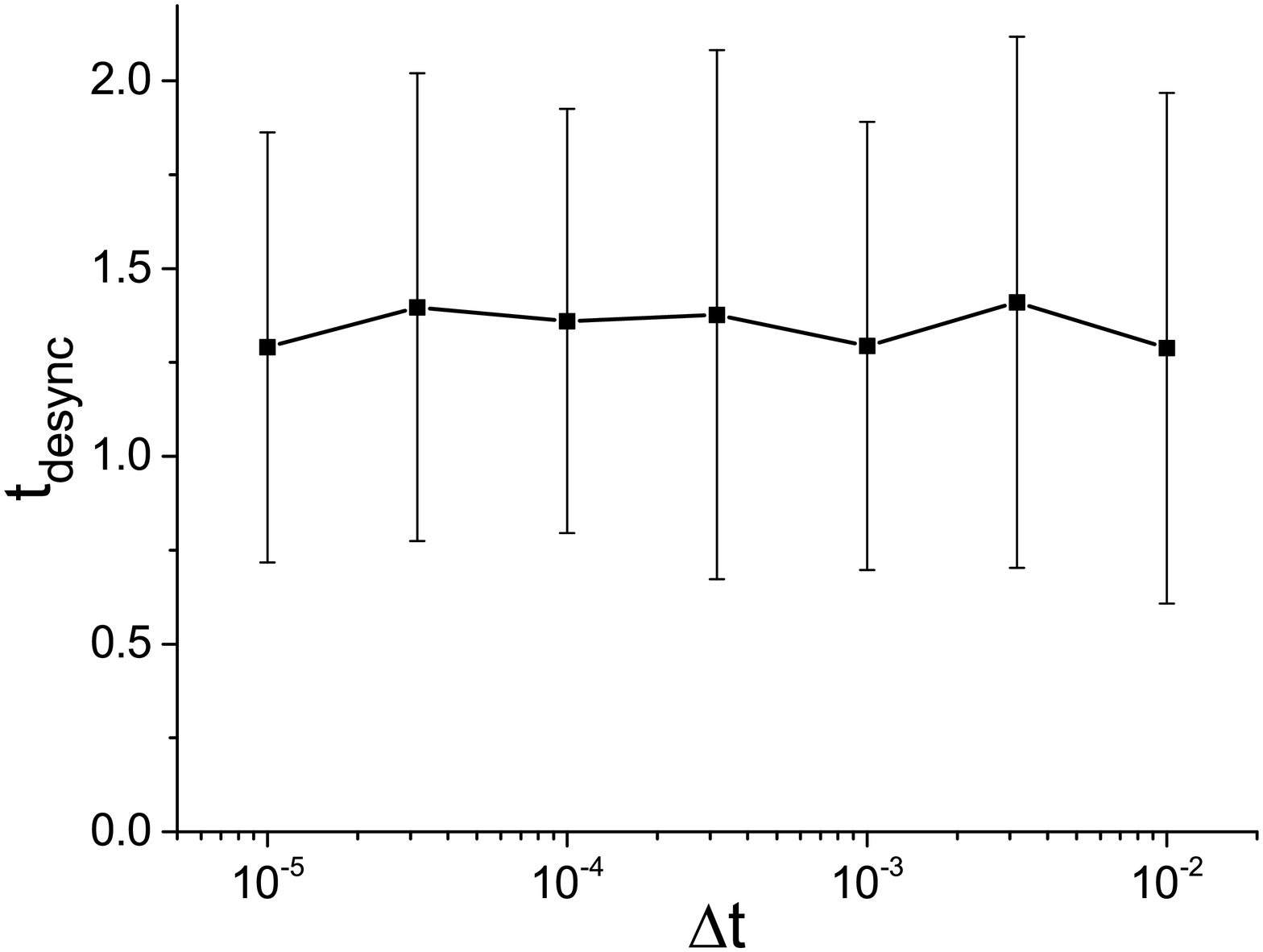}
\caption{Average time to desynchronization for generalized KPZ coupling [Eq.~(\ref{network_KPZ_dt})] at various $\Delta t$ time steps of integration in a ER network with $N$$=$$100$, $\langle k\rangle \approx 6$, $\nu = 0.2$, $q = 0.2$. Error bars represent standard deviation, sampled over $100$ realizations.}
\label{fig-numint_ER}
\end{figure}

\begin{figure}[ht]
\centering
\includegraphics[width=4in]{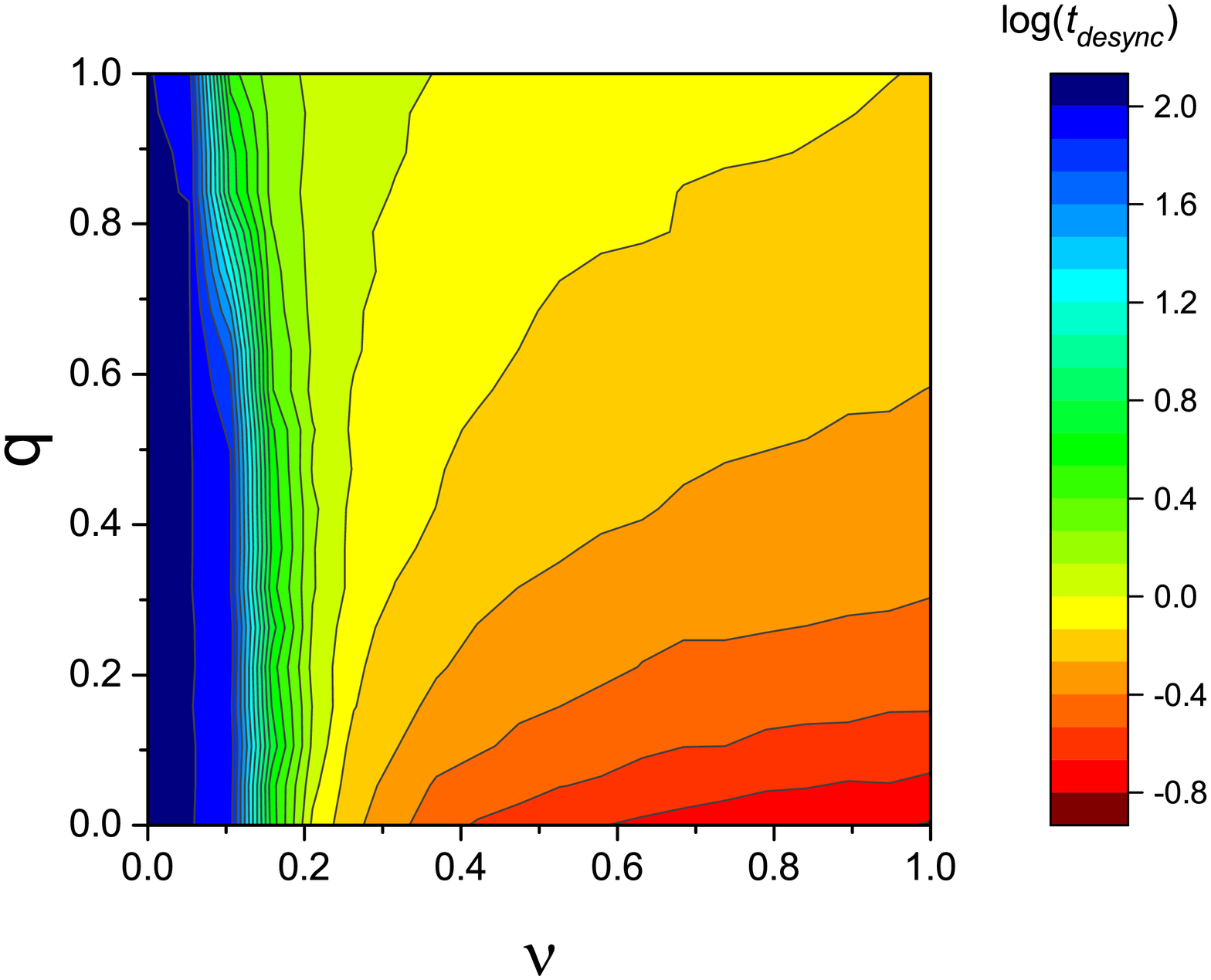}
\caption{ Time to reach desynchronization using generalized KPZ coupling [Eq.~(\ref{network_KPZ_dt})], as a function  of the delay $q = \tau / \tau_c$ and nonlinear coupling strength $\nu$, for an ER network with $N$$=$$100$ and $\langle k\rangle \approx 6$.
Data was obtained by numerically integrating Eq.~(\ref{network_KPZ_dt}) with $\Delta t=10^{-3}$.}
\label{fig-desync_ER}
\end{figure}

\begin{figure}[ht]
\centering
\includegraphics[width=\textwidth]{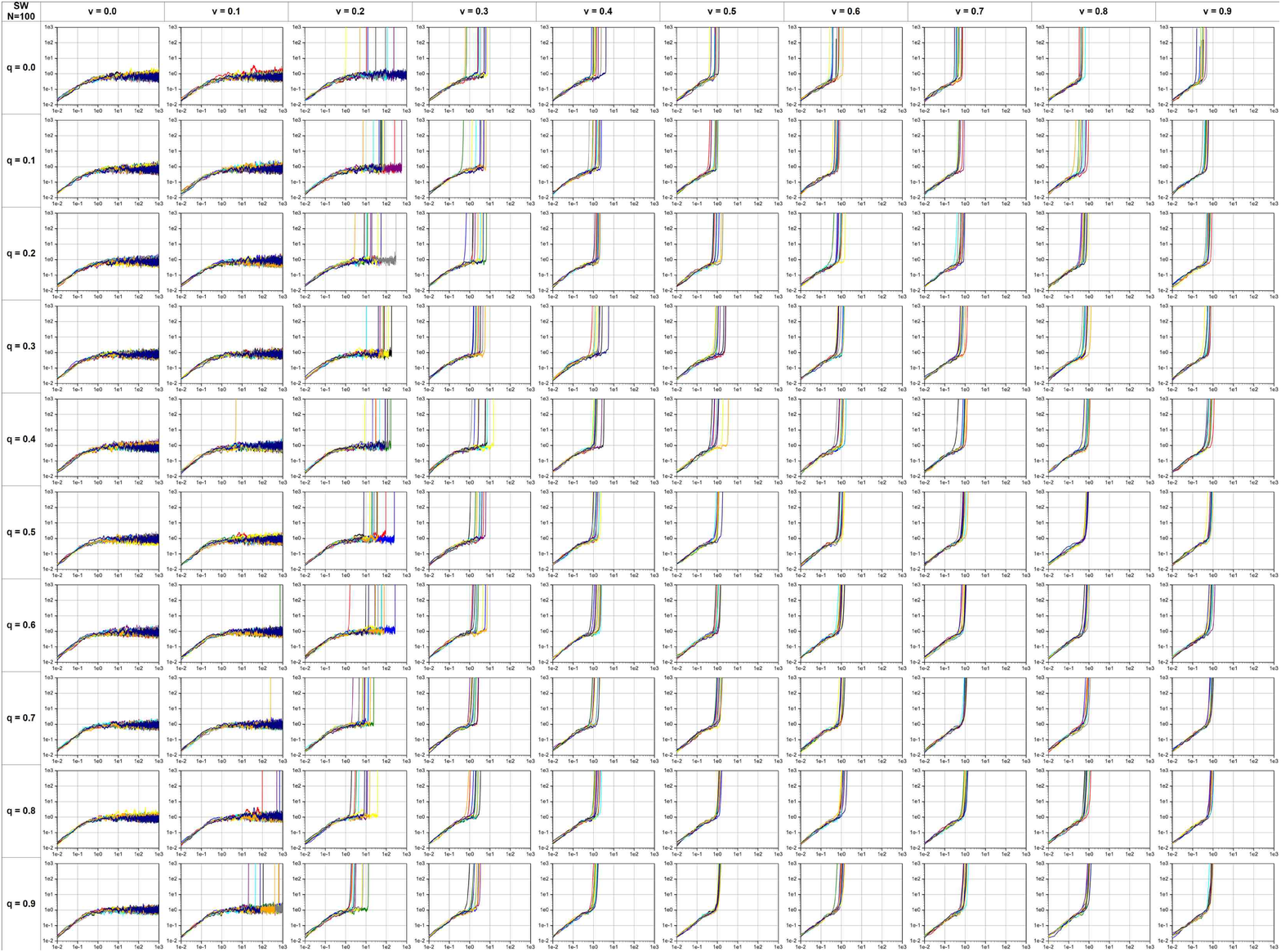}
\caption{Graphical table of the evolution of the width over time, in SW networks with $N$$=$$100$ and $\langle k\rangle \approx 3$, using generalized KPZ coupling [Eq.~(\ref{network_KPZ_dt})]. Each tile shows the width $\langle w^2 \rangle$ as a function of time, using identical scales in each tile.
Data was obtained by numerically integrating Eq.~(\ref{network_KPZ_dt}) with $\Delta t=10^{-3}$.
Colors represent $10$ distinct realizations of noise. Each row corresponds to the indicated value of time delay $q = \tau / \tau_c$, and each column corresponds to the indicated nonlinear coupling strength $\nu$. A high-resolution version of this figure is provided separately in TIFF format among the Supplemental Materials (W2vsTimeMapSW\_k3.tif).}
\label{fig-map_SW_k3}
\end{figure}

\begin{figure}[ht]
\centering
\includegraphics[width=\textwidth]{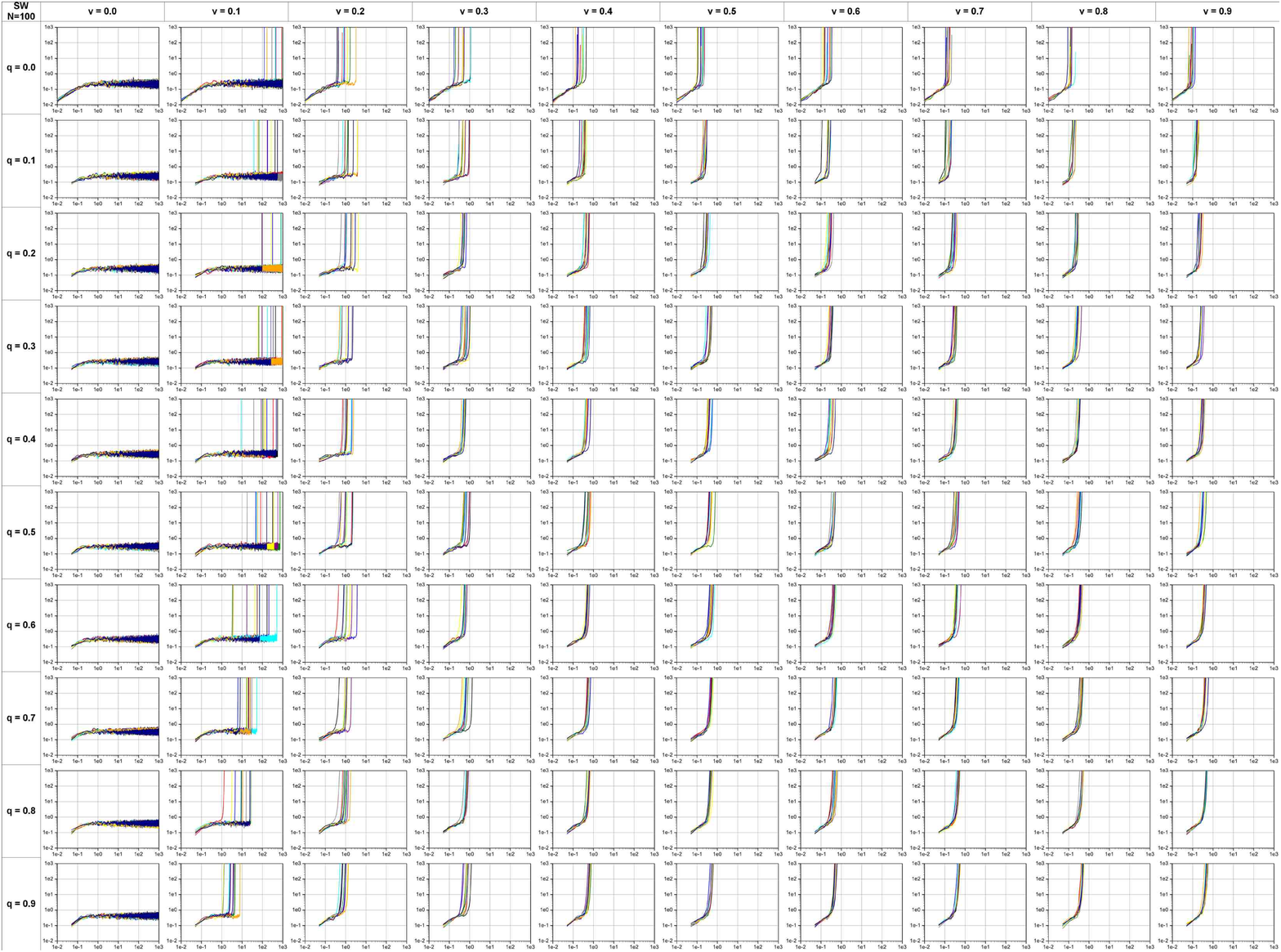}
\caption{Graphical table of the evolution of the width over time, in SW networks with $N$$=$$100$ and $\langle k\rangle \approx 6$, using generalized KPZ coupling [Eq.~(\ref{network_KPZ_dt})]. Each tile shows the width $\langle w^2 \rangle$ as a function of time, using identical scales in each tile.
Data was obtained by numerically integrating Eq.~(\ref{network_KPZ_dt}) with $\Delta t=10^{-3}$.
Colors represent $10$ distinct realizations of noise. Each row corresponds to the indicated value of time delay $q = \tau / \tau_c$, and each column corresponds to the indicated nonlinear coupling strength $\nu$. A high-resolution version of this figure is provided separately in TIFF format among the Supplemental Materials (W2vsTimeMapSW\_k6.tif).}
\label{fig-map_SW_k6}
\end{figure}

\begin{figure}[ht]
\centering
\includegraphics[width=4in]{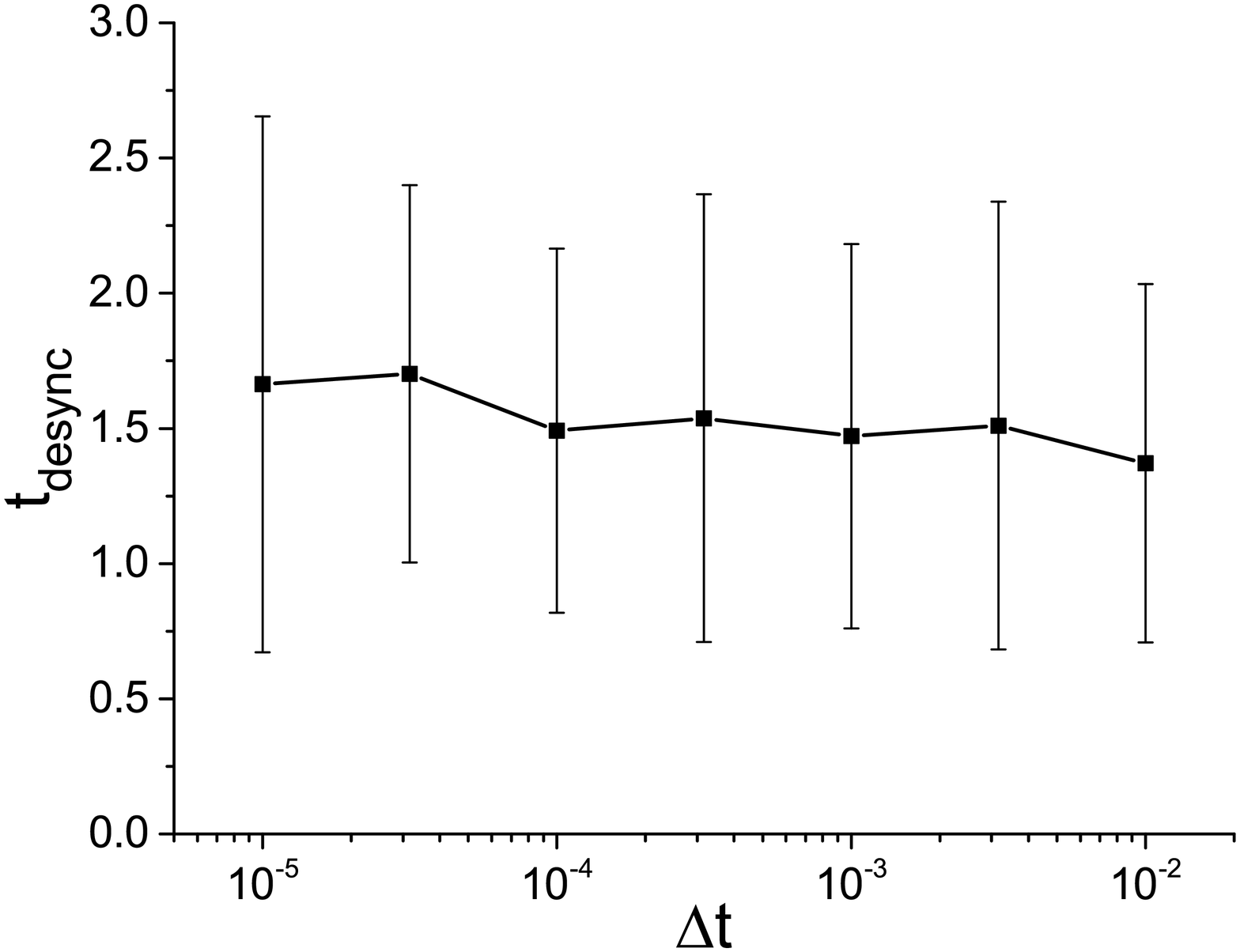}
\caption{Average time to desynchronization for generalized KPZ coupling [Eq.~(\ref{network_KPZ_dt})] at various $\Delta t$ time steps of integration in a SW network with $N$$=$$100$, $\langle k\rangle \approx 6$, $\nu = 0.2$, $q = 0.2$. Error bars represent standard deviation, sampled over $100$ realizations.}
\label{fig-numint_SW}
\end{figure}

\begin{figure}[ht]
\centering
\includegraphics[width=4in]{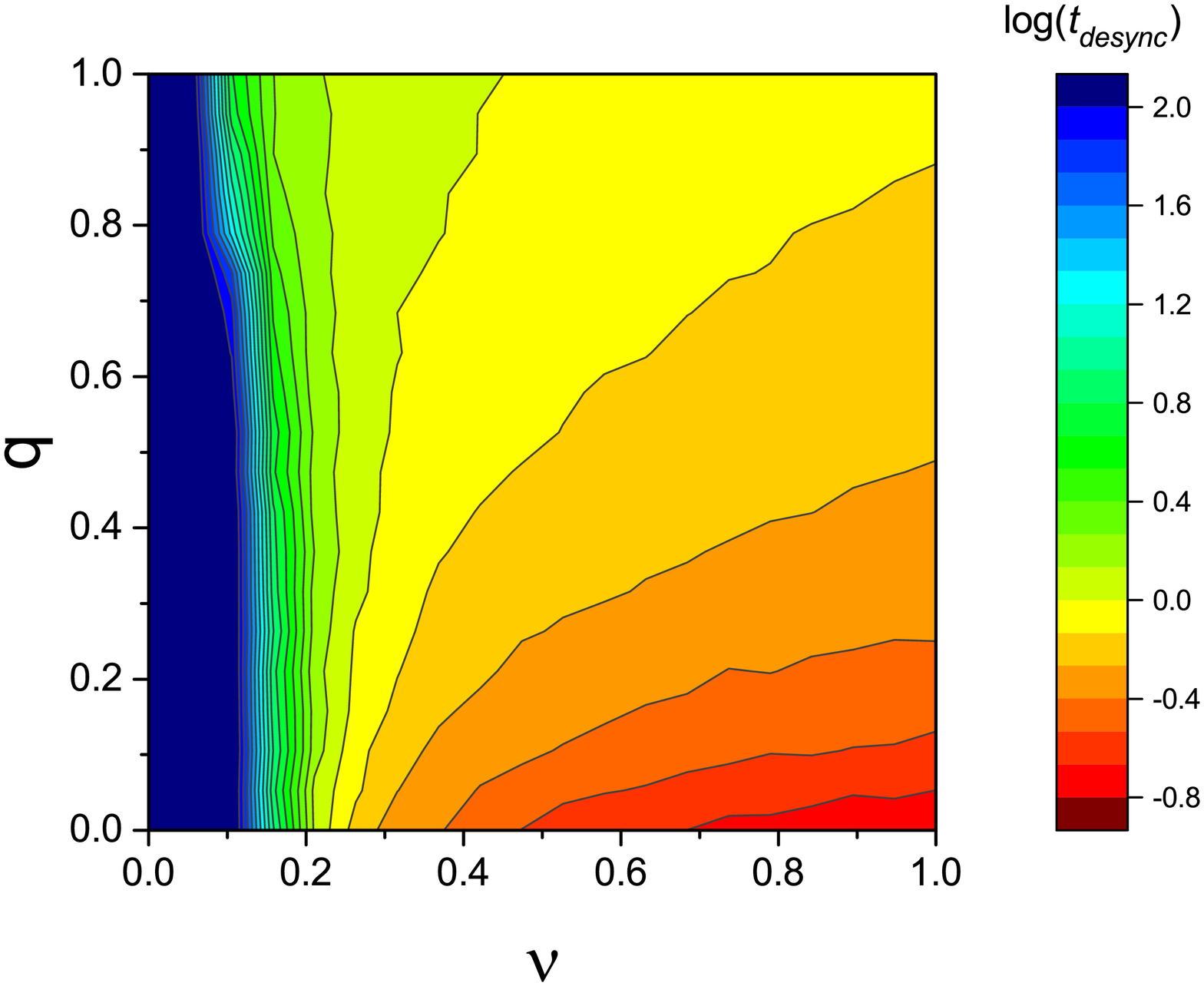}
\caption{ Time to reach desynchronization using generalized KPZ coupling [Eq.~(\ref{network_KPZ_dt})], as a function  of the delay $q = \tau / \tau_c$ and nonlinear coupling strength $\nu$, for a SW network with $N$$=$$100$ and $\langle k\rangle \approx 6$.
Data was obtained by numerically integrating Eq.~(\ref{network_KPZ_dt}) with $\Delta t=10^{-3}$.}
\label{fig-desync_SW}
\end{figure}

In this Supplemental Material, we provide some additional results (figures) obtained by the numerical integration of the stochastic delay differential equation
%%%%%%%%%%%%%%%%%%%%%%%%%%%%%%%%%%%%%%%%%%%%%%%%%%%%%%%%%%%%%%%%%%%%%%%%%%%%%%%%%%%%%%%%%%%%%%%%%%%%%%
\begin{equation}
\partial_t h_i(t) = -\sum_{j = 1}^N C_{ij}(h_i - h_j) + \nu\sum_{j,k = 1 (j<k)}^N C_{ij}C_{ik}(h_j - h_i)(h_k - h_i) + \eta_i(t) \;,
\label{SM_network_KPZ}
\end{equation}
%%%%%%%%%%%%%%%%%%%%%%%%%%%%%%%%%%%%%%%%%%%%%%%%%%%%%%%%%%%%%%%%%%%%%%%%%%%%%%%%%%%%%%%%%%%%%%%%%%%%%%
discussed in Sec.~V. in the main text. The effects of time delays are captured by replacing $\{h_{l}(t)\}_{l=1}^{N}$ by $\{h_{l}(t-\tau)\}_{l=1}^{N}$ in the right-hand side of Eq.~(\ref{SM_network_KPZ}). In the above continuous-time stochastic differential equation, the noise $\eta_i(t)$ is Gaussian with zero mean and correlations $\langle \eta_i(t)\eta_j(t')\rangle = 2D\delta_{ij}\delta(t - t')\rangle$.
In the present work, we considered unweighted graphs, i.e., $C_{ij}$ is just the adjacency matrix. The time-discretized version of Eq.~(\ref{SM_network_KPZ}) becomes [1-5]
%\cite{SM_Gardiner_1985,SM_Barabasi,SM_DasSarma_PRE1997,SM_Shin_PRE1998b,SM_Majumdar_2005}
%%%%%%%%%%%%%%%%%%%%%%%%%%%%%%%%%%%%%%%%%%%%%%%%%%%%%%%%%%%%%%%%%%%%%%%%%%%%%%%%%%%%%%%%%%%%%%%%%%%%%%
\begin{equation}
h_i(t+\Delta t) - h_i(t) = -\Delta t\sum_{j = 1}^N C_{ij}(h_i - h_j) + \Delta t\,\nu\sum_{j,k = 1 (j<k)}^N C_{ij}C_{ik}(h_j - h_i)(h_k - h_i) +
\hat{\eta}_{i}(t) \sqrt{2D\Delta t} \;,
\label{network_KPZ_dt}
\end{equation}
%%%%%%%%%%%%%%%%%%%%%%%%%%%%%%%%%%%%%%%%%%%%%%%%%%%%%%%%%%%%%%%%%%%%%%%%%%%%%%%%%%%%%%%%%%%%%%%%%%%%%%
where $\hat{\eta}_{i}(t)$ are independent and identically distributed random variables for all $i$ and $t$ with Gaussian distribution of zero mean and unit variance. In our simulations, without loss of generality, we set $D$$=$$1$. For example, in one dimension, with nearest neighbor couplings and periodic boundary conditions, the above equation becomes
%%%%%%%%%%%%%%%%%%%%%%%%%%%%%%%%%%%%%%%%%%%%%%%%%%%%%%%%%%%%%%%%%%%%%%%%%%%%%%%%%%%%%%%%%%%%%%%%%%%%%%
\begin{equation}
h_i(t+\Delta t) - h_i(t) = -\Delta t(2h_i - h_{i+1} - h_{i-1}) + \Delta t\,\nu(h_{i+1} - h_i)(h_{i-1} - h_i) +
\hat{\eta}_{i}(t) \sqrt{2D\Delta t} \;.
\label{1d_KPZ_dt}
\end{equation}
%%%%%%%%%%%%%%%%%%%%%%%%%%%%%%%%%%%%%%%%%%%%%%%%%%%%%%%%%%%%%%%%%%%%%%%%%%%%%%%%%%%%%%%%%%%%%%%%%%%%%%
Employing Eqs.~(\ref{network_KPZ_dt}) and (\ref{1d_KPZ_dt}), we explore and display the width of the synchronization landscape as a function of time for a one-dimensional lattice (with nearest-neighbor coupling and periodic boundary conditions) [Fig.~\ref{fig-map_1d}], ER graphs [Fig.~\ref{fig-map_ER}], and SW networks [Figs.~\ref{fig-map_SW_k3}, \ref{fig-map_SW_k6}]. We also demonstrate that the instability observed in the resulting stochastic landscapes is {\em not} the result of insufficient time discretization in the numerical integration scheme, but rather, it is intrinsic to the generalized KPZ coupling on discrete structures (such as lattices or networks) for sufficiently strong nonlinear coupling and/or large time delays. To that end, we show the average time to desynchronization (the onset of instability) as a function of $\Delta t$ used in the above numerical integration scheme for a one-dimensional lattice [Fig.~\ref{fig-numint_1d}], ER networks [Fig.~\ref{fig-numint_ER}], and SW networks [Fig.~\ref{fig-numint_SW}]. Finally, we show the average time to desynchronization  as a function of the nonlinear coupling $\nu$ and relative time delay $q=\tau/\tau_{c}$ in a one-dimensional lattice [Fig.~\ref{fig-desync_1d}], ER networks [Fig.~\ref{fig-desync_ER}], and SW networks [Fig.~\ref{fig-desync_SW}]. Our results for SW networks [Figs.~\ref{fig-map_SW_k3} and \ref{fig-map_SW_k6}] also suggest that while increasing connectivity (average degree $\langle k\rangle$) reduces the width in the quasi-stationary state, the quasi-stationary period is becoming shorter, i.e., the system is more prone to instability.

%%%%%%%%%%%%%%%%%%%%%%%%%%%%%%%%%%%%%%%%%%%%%%%%%%%%%%%%%%%%%%%%%%%%%%%%%%%%%%%%%%%%%%%%%%%%%%%%%%%%%%
%%%%%%%%%%%%%%%%%%%%%%%%%%%%%%%%%%%%  1d plots with KPZ nonlinearity %%%%%%%%%%%%%%%%%%%%%%%%%%%%%%%%%
%%%%%%%%%%%%%%%%%%%%%%%%%%%%%%%%%%%%%%%%%%%%%%%%%%%%%%%%%%%%%%%%%%%%%%%%%%%%%%%%%%%%%%%%%%%%%%%%%%%%%%
%%%%%%%%%%%%%%%%%%%%%%%%%%%%%%%%%%%%%%%%%%%%%%%%%%%%%%%%%%%%%%%%%%%%%%%%%%%%%%%%%%%%%%%%%%%%%%%%%%%%%%%%

%%%%%%%%%%%%%%%%%%%%%%%%%%%%%%%%%%%%%%%%%%%%%%%%%%%%%%%%%%%%%%%%%%%%%%%%%%%%%%%%%%%%%%%%%%%%%%%%%%%%%
%%%%%%%%%%%%%%%%%%%%%%%%%%%%%%%%%%%%%%%%%%%%%%%%%%%%%%%%%%%%%%%%%%%%%%%%%%%%%%%%%%%%%%%%%%%%%%%%%%%%%%

%%%%%%%%%%%%%%%%%%%%%%%%%%%%%%%%%%%%%%%%%%%%%%%%%%%%%%%%%%%%%%%%%%%%%%%%%%%%%%%%%%%%%%%%%%%%%%%%%%%%%%
%%%%%%%%%%%%%%%%%%%%%%%%%%%%%%%%%%%%%%%%%%%%%%%%%%%%%%%%%%%%%%%%%%%%%%%%%%%%%%%%%%%%%%%%%%%%%%%%%%%%%%

%%%%%%%%%%%%%%%%%%%%%%%%%%%%%%%%%%%%%%%%%%%%%%%%%%%%%%%%%%%%%%%%%%%%%%%%%%%%%%%%%%%%%%%%%%%%%%%%%%%%%%
%%%%%%%%%%%%%%%%%%%%%%%%%%%%%%%%%%%%  ER plots with KPZ nonlinearity %%%%%%%%%%%%%%%%%%%%%%%%%%%%%%%%%
%%%%%%%%%%%%%%%%%%%%%%%%%%%%%%%%%%%%%%%%%%%%%%%%%%%%%%%%%%%%%%%%%%%%%%%%%%%%%%%%%%%%%%%%%%%%%%%%%%%%%%
%%%%%%%%%%%%%%%%%%%%%%%%%%%%%%%%%%%%%%%%%%%%%%%%%%%%%%%%%%%%%%%%%%%%%%%%%%%%%%%%%%%%%%%%%%%%%%%%%%%%%%%%

%%%%%%%%%%%%%%%%%%%%%%%%%%%%%%%%%%%%%%%%%%%%%%%%%%%%%%%%%%%%%%%%%%%%%%%%%%%%%%%%%%%%%%%%%%%%%%%%%%%%%%
%%%%%%%%%%%%%%%%%%%%%%%%%%%%%%%%%%%%%%%%%%%%%%%%%%%%%%%%%%%%%%%%%%%%%%%%%%%%%%%%%%%%%%%%%%%%%%%%%%%%%%

%%%%%%%%%%%%%%%%%%%%%%%%%%%%%%%%%%%%%%%%%%%%%%%%%%%%%%%%%%%%%%%%%%%%%%%%%%%%%%%%%%%%%%%%%%%%%%%%%%%%%%
%%%%%%%%%%%%%%%%%%%%%%%%%%%%%%%%%%%%%%%%%%%%%%%%%%%%%%%%%%%%%%%%%%%%%%%%%%%%%%%%%%%%%%%%%%%%%%%%%%%%%%

%%%%%%%%%%%%%%%%%%%%%%%%%%%%%%%%%%%%%%%%%%%%%%%%%%%%%%%%%%%%%%%%%%%%%%%%%%%%%%%%%%%%%%%%%%%%%%%%%%%%%%
%%%%%%%%%%%%%%%%%%%%%%%%%%%%%%%%%%%%  SW plots with KPZ nonlinearity %%%%%%%%%%%%%%%%%%%%%%%%%%%%%%%%%

%%%%%%%%%%%%%%%%%%%%%%%%%%%%%%%%%%%%%%%%%%%%%%%%%%%%%%%%%%%%%%%%%%%%%%%%%%%%%%%%%%%%%%%%%%%%%%%%%%%%%%
%%%%%%%%%%%%%%%%%%%%%%%%%%%%%%%%%%%%%%%%%%%%%%%%%%%%%%%%%%%%%%%%%%%%%%%%%%%%%%%%%%%%%%%%%%%%%%%%%%%%%%%%

%%%%%%%%%%%%%%%%%%%%%%%%%%%%%%%%%%%%%%%%%%%%%%%%%%%%%%%%%%%%%%%%%%%%%%%%%%%%%%%%%%%%%%%%%%%%%%%%%%%%%%
%%%%%%%%%%%%%%%%%%%%%%%%%%%%%%%%%%%%%%%%%%%%%%%%%%%%%%%%%%%%%%%%%%%%%%%%%%%%%%%%%%%%%%%%%%%%%%%%%%%%%%%%

%%%%%%%%%%%%%%%%%%%%%%%%%%%%%%%%%%%%%%%%%%%%%%%%%%%%%%%%%%%%%%%%%%%%%%%%%%%%%%%%%%%%%%%%%%%%%%%%%%%%%%
%%%%%%%%%%%%%%%%%%%%%%%%%%%%%%%%%%%%%%%%%%%%%%%%%%%%%%%%%%%%%%%%%%%%%%%%%%%%%%%%%%%%%%%%%%%%%%%%%%%%%%

%%%%%%%%%%%%%%%%%%%%%%%%%%%%%%%%%%%%%%%%%%%%%%%%%%%%%%%%%%%%%%%%%%%%%%%%%%%%%%%%%%%%%%%%%%%%%%%%%%%%%%
%%%%%%%%%%%%%%%%%%%%%%%%%%%%%%%%%%%%%%%%%%%%%%%%%%%%%%%%%%%%%%%%%%%%%%%%%%%%%%%%%%%%%%%%%%%%%%%%%%%%%%

%\begin{figure}[ht]
%\centering
%\includegraphics[width=\textwidth]{PDmax_ab.eps}
%\caption{Distribution of the maximum using KPZ coupling in ER networks. (a) unscaled distribution, (b) scaled distribution. Parameters: $N$$=$$100$, $\langle k\rangle \approx 6$, $\nu = 0.1$. Samples collected in time interval $10 < t <50$ from $50$ distinct realizations of noise ($4000$ samples each) that did not desynchronize until $t=50$. }
%\label{fig-pdmax}
%\end{figure}
%
%\begin{figure}[ht]
%\centering
%\includegraphics[width=\textwidth]{PDmin_ab.eps}
%\caption{Distribution of the minimum using KPZ coupling in ER networks. (a) unscaled distribution, (b) scaled distribution. Parameters: $N$$=$$100$, $\langle k\rangle \approx 6$, $\nu = 0.1$. Samples collected in time interval $10 < t <50$ from $50$ distinct realizations of noise ($4000$ samples each) that did not desynchronize until $t=50$.}
%\label{fig-pdmin}
%\end{figure}

\vspace*{1cm}

%\begin{thebibliography}{99}

\small
%\bibitem{SM_Gardiner_1985}
\noindent
[1] C.W. Gardiner, {\it Handbook of Stochastic Methods}, 2nd ed. (Springer-Verlag, New York, 1985).

%\bibitem{SM_Barabasi}
\noindent
[2] A.-L. Barab\'asi and H.E. Stanley,
{\it Fractal Concepts in Surface Growth}
(Cambridge University Press, Cambridge, 1995).

%\bibitem{SM_DasSarma_PRE1997}
\noindent
[3] C. Dasgupta, J.M. Kim, M. Dutta, and S. Das Sarma,
%``Instability, intermittency, and multiscaling in discrete growth models of kinetic roughening",
Phys. Rev. E {\bf 55}, 2235 (1997).

%\bibitem{SM_Shin_PRE1998b}
\noindent
[4] C.-H. Lam and F.G. Shin,
%``Improved discretization of the Kardar-Parisi-Zhang equation",
Phys. Rev. E {\bf 58}, 5592 (1998).

%\bibitem{SM_Majumdar_2005}
\noindent
[5] S.N. Majumdar and A. Comtet,
%``Airy Distribution Function: From the Area Under a Brownian Excursion to the Maximal Height of Fluctuating Interfaces''
J. Stat. Phys. {\bf 119}, 777 (2005).

%\end{thebibliography}


\clearpage
\begin{thebibliography}{99}

\bibitem{Saber_IEEE2007}
R. Olfati-Saber, J.A. Fax, and R.M. Murray,
%``Consensus and cooperation in networked multi-agent systems,"
Proc. IEEE {\bf 95}, 215 (2007).

\bibitem{Arenas_PhysRep2008}
A. Arenas {\it et al.},
%``Synchronization in complex networks,"
Phys. Rep. {\bf 469}, 93 (2008).

\bibitem{Korniss_Springer2012}
G. Korniss, R. Huang, S. Sreenivasan, and B.K. Szymanski,
%``Optimizing synchronization, flow, and robustness in weighted complex networks,"
in {\it Handbook of Optimization in Complex Networks: Communication and Social Networks}, edited by M.T. Thai and P. Pardalos, Springer
Optimization and Its Applications Vol. 58, (Springer, New York, 2012) pp. 61-96.

\bibitem{Sipahi_IEEE2011}
R. Sipahi {\it et al},
%``Stability and stabilization of systems with time delay,"
IEEE Contr. Sys. {\bf 31}, 38 (2011).

\bibitem{HuntPRL}
D. Hunt, G. Korniss, and B.K. Szymanski,
%``Network synchronization in a noisy environment with time delays: fundamental limits and trade-Offs,"
Phys. Rev. Lett. {\bf 105}, 068701 (2010).

\bibitem{HuntPLA}
D. Hunt, G. Korniss, and B.K. Szymanski,
%``The impact of competing time delays in coupled stochastic systems,"
Phys. Lett. A {\bf 375}, 880 (2011).

\bibitem{HuntPRE}
D. Hunt, B.K. Szymanski, and G. Korniss,
%``Network coordination and synchronization in a noisy environment with time delays,''
Phys. Rev. E {\bf 86}, 056114 (2012).

\bibitem{Hod_PRL2010}
S. Hod,
%``Analytic treatment of the network synchronization problem with time delays,"
Phys. Rev. Lett. {\bf 105}, 208701 (2010).

\bibitem{Chen_EPL2008}
Q. Wang, Z. Duan, M. Perc, and G. Chen,
%``Synchronization transitions on small-world neuronal networks: effects of information transmission delay and rewiring probability,"
Europhys. Lett. {\bf 83}, 50008 (2008).

\bibitem{Chen_PRE2009}
Q. Wang, M. Perc, Z. Duan, and G. Chen,
%``Synchronization transitions on a scale-free neuronal networks due to finite information transmission delays,"
Phys. Rev. E {\bf 80}, 026206 (2009).

\bibitem{Chen_PLOS2011}
Q. Wang, G. Chen, and M. Perc,
%``Synchronous bursts on scale-free neuronal networks with attractive and repulsive coupling,"
PLoS One {\bf 6}, e15851 (2011).

\bibitem{Kalecki_1935}
M. Kalecki,
%``A macrodynamic theory of business cycles,"
Econometrica {\bf 3}, 327 (1935).

\bibitem{Frisch_1935}
R. Frisch and H. Holme,
%``The characteristic solutions of a mixed difference and differential equation occuring in economic dynamics,"
Econometrica {\bf 3}, 225 (1935).

\bibitem{Kuechler_SSR1992}
U. K\"uchler and B. Mensch,
%``Langevin stochastic differential equation extended by a time-delayed term,"
Stoch. and Stoch. Rep. {\bf 40}, 23 (1992).

\bibitem{Ohira_PRE2000}
T. Ohira and T. Yamane,
%``Delayed stochastic systems,"
Phys. Rev. E. {\bf 61}, 1247 (2000).

\bibitem{Frank_PRE2001}
T. D. Frank and P.J. Beek,
%``Stationary solutions of linear stochastic delay-differential equations: aplications to biological syatems,"
Phys. Rev. E. {\bf 64}, 021917 (2001).

\bibitem{Huberman_IEEE1991}
T. Hogg and B.A. Huberman,
%``Controlling chaos in distributed systems,"
IEEE Trans. on Sys., Man, and Cybernetics
{\bf 21}, 1325 (1991).

\bibitem{Strogatz_PRE2003}
M.G. Earl and S.H. Strogatz,
%``Synchronization in oscillator networks with delayed coupling: A stability criterion,"
Phys. Rev. E {\bf 67}, 036204 (2003).

\bibitem{Barahona_PRL2002}
M. Barahona and L.M. Pecora,
%``Synchronization in small-world systems,"
Phys. Rev. Lett. {\bf 89}, 054101 (2002).

\bibitem{Nishikawa_PRL2002}
T. Nishikawa, A.E. Motter, Y.-C. Lai, and F.C. Hoppensteadt,
%``Heterogeneity in oscillator networks: are smaller worlds easier to synchronize?,"
Phys. Rev. Lett. {\bf 91}, 014101 (2003).

\bibitem{LaRocca_PRE2008}
C.E. La Rocca, L.A. Braunstein, and P.A. Macri,
%``Evolution equation for a model of surface relaxation in complex networks,"
Phys. Rev. E {\bf 77}, 046120 (2008).

\bibitem{LaRocca_PRE2009}
C.E. La Rocca, L.A. Braunstein, and P.A. Macri,
%``Conservative model for synchronization problems in complex networks,''
Phys. Rev. E {\bf 80}, 026111 (2009).

\bibitem{Zhou_PRL2006}
C. Zhou, A.E. Motter, and J. Kurths,
%``Universality in the synchronization of weighted random networks,"
Phys. Rev. Lett. {\bf 96}, 034101 (2006).

\bibitem{GK_PRE2007}
G. Korniss,
%``Synchronization in weighted uncorrelated complex networks in a noisy environment: optimization and connections with transport efficiency,"
Phys. Rev. E {\bf 75}, 051121 (2007).

\bibitem{Lai_Chaos2008}
W.-X. Wang, L. Huang, Y.-C. Lai, and G.-R. Chen,
%``Onset of synchronization in weighted scale-free networks",
Chaos {\bf 19}, 013134 (2009).

\bibitem{Saavedra_PNAS2012}
S. Saavedra, K. Hagerty, and B. Uzzi,
%``Synchronicity, instant messaging and performance among financial traders,"
Proc. Natl. Acad. Sci. U.S.A. {\bf 108}, 5296 (2011).

\bibitem{fireflies}
H. M. Smith,
%``Synchronous flashing of fireflies,"
Science {\bf 82}, 151 (1935).

\bibitem{Reynolds_CG1987}
C. G. Reynolds
%``Flocks, herds, and schools: a distributed behavioral model,"
{\it Computer Graphics} {\bf 21}, 25--34 (1987).

\bibitem{Cucker_IEEE2007}
F. Cucker and S. Smale,
%``Emergent behavior in flocks,"
IEEE Trans. Automat. Contr. {\bf 52}, 852 (2007).

\bibitem{Vicsek_PRL1995}
T. Vicsek, A. Czir\'ok, E. Ben-Jacob, I. Cohen, and O. Shochet,
%``Novel type of phase transition in a system of self-driven particles,"
Phys. Rev. Lett. {\bf 75}, 1226 (1995).

\bibitem{Hutchinson_1948}
G.E. Hutchinson,
%``Circular causal systems in ecology,"
Ann. N.Y. Acad. Sci. {\bf 50}, 221 (1948).

\bibitem{May_1973}
R.M. May,
%``Time-delay versus stability in population models with two and three trophic levels,"
Ecology {\bf 54}, 315--325 (1973).

\bibitem{Ruan_2006}
S. Ruan,
%``Delay differential equations in single species dynamics,"
in {\it Delay Differential Equations and Applications}, edited by O. Arino, M.L. Hbid, and E.A. Dads, NATO Science Series II: Mathematics, Physics and Chemistry, Vol. 205 (Springer, Berlin, 2006) pp.~477--517.

\bibitem{Milton_EPL2008}
J.G. Milton, J.L. Cabrera, and T. Ohira,
%``Unstable dynamical systems: delays, noise and control,"
Europhys. Lett. {\bf 83}, 48001 (2008).

\bibitem{Milton_PTRSA2009}
J. Milton, J.L. Townsend, M.A. King, and T. Ohira,
%``Balancing with positive feedback: the case for discontinuous control,"
Phil. Trans. R. Soc. A {\bf 367}, 1181 (2009).

\bibitem{Cabrera_PRL2002}
J.L. Cabrera and J.G. Milton,
%``On-off intermittency in a human balancing task,"
Phys. Rev. Lett. {\bf 89}, 158702 (2002).

\bibitem{Cabrera_CMP2006}
J.L. Cabrera, C. Luciani, and J. Milton,
%``Neural control on multiple time scales: insights form human stick balancing,"
Cond. Matt. Phys. {\bf 9}, 373 (2006).

\bibitem{Izhikevich_SIAM2001}
E. Izhikevich,
%``Synchronization of elliptic bursters,"
SIAM Rev. {\bf 43}, 315 (2001).

\bibitem{Johari_IEEE2001}
R. Johari and D. Kim Hong Tan,
%``End-to-end congestion control for the internet: delays and stability,"
IEEE/ACM Trans. Networking {\bf 9}, 818 (2001).

\bibitem{Saber_IEEE2004}
R. Olfati-Saber and R.M. Murray,
%``Consensus problems in networks of agents with switching topology and time-delays,"
IEEE Trans. Automat. Contr. {\bf 49}, 1520 (2004).

\bibitem{Ott_2006}
T.J. Ott,
``On the Ornstein-Uhlenbeck process with delayed feedback," \\
http://www.teunisott.com/Papers/TCP\_Paradigm/Del\_O\_U.pdf (2006), Date Last Accessed 06/05/2015.

\bibitem{GK_Science2003}
G. Korniss, M.A. Novotny, H. Guclu, Z. Toroczkai, and P.A. Rikvold,
%``Suppressing roughness of virtual times in parallel discrete-event simulations,"
Science {\bf 299}, 677 (2003).

\bibitem{Korniss_PRL2000}
G. Korniss, Z. Toroczkai, M.A. Novotny, and P.A. Rikvold,
%``From massively parallel algorithms and fluctuating time horizons to non-equilibrium surface growth,"
Phys. Rev. Lett. {\bf 84}, 1351 (2000).

\bibitem{Guclu_PRE2006}
H. Guclu, G. Korniss, M.A. Novotny, Z. Toroczkai, and Z. R\'acz,
%``Synchronization landscapes in small-world-connected computer networks,"
Phys. Rev. E {\bf 73}, 066115 (2006).

\bibitem{Guclu_Chaos2007}
H. Guclu, G. Korniss, and Z. Toroczkai,
%``Extreme fluctuations in noisy task-completion landscapes on scale-free networks,"
Chaos {\bf 17} 026104 (2007).

\bibitem{Orosz_PRSA2006}
G. Orosz and G. Stepan,
%``Subcritical Hopf bifurcations in a car-following model with reaction-time delay,"
Proc. R. Soc. A {\bf 462}, 2643 (2006).

\bibitem{Orosz_PTRSA2010}
G. Orosz, R.E. Wilson, and G. Stepan,
%``Traffic jams: dynamics and control,"
Phil. Trans. R. Soc. A {\bf 368}, 4455 (2010).

\bibitem{Fax_IEEE2004}
J.A. Fax and R.M. Murray,
%``Information flow and cooperative control of vehicle formations,"
IEEE Trans. Automat. Contr. {\bf 49}, 1465 (2004).

%%%%%%  complex networks %%%%%%%%%%%%%%%%%%%%%%%%%%%%%%%%%%%%%%%%%%%%%%%%%%%

\bibitem{Watts_Nature1998}
D.J. Watts and S.H. Strogatz,
Nature {\bf 393}, 440 (1998).

\bibitem{Barab_sci}
A.-L. Barab\'asi and R. Albert,
Science {\bf 286}, 509 (1999).

\bibitem{BarabREV}
R. Albert and A.-L. Barab\'asi,
Rev. Mod. Phys. {\bf 74}, 47 (2002).

\bibitem{MendesREV}
S.N. Dorogovtsev and J.F.F. Mendes,
Adv. in Phys. {\bf 51}, 1079 (2002).

\bibitem{ER_1960}
P. Erd\H{o}s and A. R\'enyi,
%``On the evolution of random graphs,"
Publ. Math. Inst. Hung. Acad. Sci. {\bf 5} 17 (1960).

%%%%%%%%%%%%%%%%%%%%%%%%%%%%%%%%%%%%%%%%%%%%%%%%%%%%%%%%%%%%%%%%%%%%%%%%%%%%%%%%%%%%%%%%%%%%%

\bibitem{Guclu2004}
H. Guclu and G. Korniss,
%``Extreme fluctuations in small-worlds with relaxational dynamics,"
Phys. Rev. E {\bf 69}, 065104(R) (2004).

\bibitem{Guclu_FNL}
H. Guclu and G. Korniss,
%``Extreme fluctuations in small-world-coupled autonomous systems with relaxational dynamics,"
Fluct. Noise Lett. {\bf 5}, L43 (2005).

\bibitem{Lai_SREP2014}
Y.-Z. Chen, Z.-G. Huang, and Y.-C. Lai,
%``Controlling extreme events on complex networks",
Sci. Rep. {\bf 4}, 6121 (2014).

\bibitem{Majumdar_2004}
S.N. Majumdar and A. Comtet,
%``Exact Maximal Height Distribution of Fluctuating Interfaces''
Phys. Rev. Lett. {\bf 92}, 225501 (2004).

\bibitem{Majumdar_2005}
S.N. Majumdar and A. Comtet,
%``Airy Distribution Function: From the Area Under a Brownian Excursion to the Maximal Height of Fluctuating Interfaces''
J. Stat. Phys. {\bf 119}, 777 (2005).

\bibitem{Raychaudhuri_PRL2001}
S. Raychaudhuri, M. Cranston, C. Przybyla, and Y. Shapir,
%"Maximal height scaling of kinetically growing surfaces",
Phys. Rev. Lett. {\bf 87}, 136101 (2001).

\bibitem{FORWZ_PRE1994}
G. Foltin, K. Oerding, Z. Racz, R. L. Workman, and R. K. P. Zia,
%"Width distribution for random-walk interfaces",
Phys. Rev. E {\bf 50}, R639 (1994).

\bibitem{Fisher1928}
R.A. Fisher and L.H.C. Tippett,
%``The frequency distribution of the largest or smallest member of a sample,"
Proc. Camb. Philos. Soc. {\bf 24}, 180--191 (1928).

\bibitem{Gumbel1958}
E.J. Gumbel,
{\it Statistics of Extremes},
(New York, Columbia University Press, 1958).

\bibitem{Galambos1994}
{\it Extreme Value Theory and Applications},
eds. J. Galambos, J. Lechner, and E. Simin,
(Kluwer Academic Publishers, Dordrecht 1994).

%%%%%%%%%%%%%%%%%%%%%%%%%%%%%%%%%%%%%%%%%%%%%%%%%%%%%%%%%%%%%%%%%%%%%%%%%%%%%%%%%%%%%%%%%%%
\bibitem{exp_tail}
For example, for eponential-like asymptotic tail behavior, $P_{>}(x) \simeq e^{-cx^{\delta}}$,
$a_N= \left(\frac{\ln N}{c}\right)^{1/\delta}$ and
$b_N = (\delta c)^{-1}\left(\frac{\ln N}{c}\right)^{(1/\delta) - 1}$ \cite{Fisher1928,Gumbel1958,Galambos1994}.
%%%%%%%%%%%%%%%%%%%%%%%%%%%%%%%%%%%%%%%%%%%%%%%%%%%%%%%%%%%%%%%%%%%%%%%%%%%%%%%%%%%%%%%%%%%%

\bibitem{OMalley_2008}
L. O'Malley, G. Korniss, and T. Caraco,
%``Ecological Invasion, Roughened Fronts, and a Competitor's Extreme Advance: Integrating Stochastic Spatial-Growth Models''
Bull. Math. Biol. {\bf 71}, 1160 (2009).

\bibitem{Hayes_1950}
N.D. Hayes,
%``Roots of the transcendental equation associated with a certain difference-differential equation,"
J. London Math. Soc. {\bf s1-25}, 226 (1950).

%%%%%%%%%%%%%%%%%%%%%%%%%%%%%%%%%%%%%%%%%%%%%%%%%%%%%%%%%%%%%%%%%%%%%%%%%%%%%%%%%%%%

\bibitem{Fiedler_1973}
M. Fiedler,
%``Algebraic connectivity of graphs,"
Czech. Math. J. {\bf 23}, 298 (1973).

\bibitem{Anderson_1985}
W.N. Anderson and T.D. Morley,
%``Eigenvalues of the Laplacian of a graph,"
Lin. Multilin. Algebra {\bf 18}, 141 (1985).

\bibitem{Boguna_EPJB2004}
M. Bogu\~na, R. Pastor-Satorras, and A. Vespignani,
%``Cut-offs and finite-size effects in scale-free networks,"
Eur. Phys. J. B {\bf 38}, 205 (2004).

\bibitem{Catanzaro_PRE2005}
M. Catanzaro, M. Bogu\~na, and R. Pastor-Satorras,
%``Generation of uncorrelated random scale-free networks,"
Phys. Rev. E {\bf 71}, 027103 (2005).



%%%%%%%%%%%%% KPZ references %%%%%%%%%%%%%%%%%%%%%%%%%%%%%%%%%%%%%%%%%%%%%%%%%%%

\bibitem{Guclu_unpublished2008}
H. Guclu, G. Korniss, C.J. Olson Reichhardt, C. Reichhardt, and Z. Toroczkai,
unpublished (2008).

\bibitem{KPZ}
M. Kardar, G. Parisi, and Y.-C. Zhang,
Phys. Rev. Lett. {\bf 56}, 889 (1986).

\bibitem{Barabasi}
A.-L. Barab\'asi and H.E. Stanley,
{\it Fractal Concepts in Surface Growth}
(Cambridge University Press, Cambridge, 1995).

\bibitem{EW}
S.F. Edwards and D.R. Wilkinson,
Proc. R. Soc. London, Ser A {\bf 381}, 17 (1982).

\bibitem{DasSarma_PRE1996}
C. Dasgupta, S. Das Sarma, and J.M. Kim,
%``Controlled instability and multiscaling in models of epitaxial growth",
Phys. Rev. E {\bf 54}, R4552(R) (1996).

\bibitem{DasSarma_PRE1997}
C. Dasgupta, J.M. Kim, M. Dutta, and S. Das Sarma,
%``Instability, intermittency, and multiscaling in discrete growth models of kinetic roughening",
Phys. Rev. E {\bf 55}, 2235 (1997).

\bibitem{Newman_JPA1996}
T.J. Newman and A.J. Bray,
%``Strong-coupling behaviour in discrete Kardar - Parisi - Zhang equations",
J. Phys. A: Math. Gen. {\bf 29}, 7917 (1996).

\bibitem{Shin_PRE1998a}
C.-H. Lam and F.G. Shin,
%``Anomaly in numerical integrations of the Kardar-Parisi-Zhang equation",
Phys. Rev. E {\bf 57}, 6506 (1998).

\bibitem{Shin_PRE1998b}
C.-H. Lam and F.G. Shin,
%``Improved discretization of the Kardar-Parisi-Zhang equation",
Phys. Rev. E {\bf 58}, 5592 (1998).



%\bibitem{Meakin_1986}
%P. Meakin, P. Ramanlal, L. M. Sander, and R.C. Ball,
%Phys. Rev. A {\bf 34}, 5091 (1986).
%
%\bibitem{Plischke_1987}
%M. Plischke, Z. R\'acz, and D. Liu, Phys. Rev. B {\bf 35}, 3485 (1987).


%%%%%%%%%%%%%%%%%%%%%%%%%%%%%%%%
%    NOT YET CITED
%%%%%%%%%%%%%%%%%%%%%%%%%%%%%%%%

%\bibitem{Nishikawa_PRE2006}
%T. Nishikawa and A.E. Motter,
%%``Synchronization is optimal in nondiagonalizable networks,"
%Phys. Rev. E {\bf 73}, 065106(R) (2006).

%\bibitem{Nishikawa_2009}
%T. Nishikawa and A.E. Motter,
%%``Network synchronization landscape reveals compensatory structures, quantization, and the positive effect of negative interactions,"
%Proc. Natl. Acad. Sci. U.S.A.  {\bf 107}, 10342 (2010).

%\bibitem{PhysRevE.65.026139}
%H. Hong, M. Y. Choi, B. J. Kim,
%%``Synchronization on small-world networks,"
%Phys. Rev. E {\bf 65}, 026139 (2002).

%\bibitem{Kuramoto}
%Y. Kuramoto,
%in {\it Proceedings of the International Symposium on Mathematical Problems in Theoretical Physics}, edited by H. Araki,
%Lecture Notes in Physics Vol. 30  (Springer, New York, 1975) p. 420.

%\bibitem{PhysRevE.83.046206}
%G. Michael, Z. Jabeen, B. Chakraborty,
%%``Phase and frequency entrainment in locally coupled phase oscillators with repulsive interactions,"
%Phys. Rev. E {\bf 83}, 046206 (2011).

%\bibitem{optomechanical}
%G. Heinrich {\it et al.},
%%``Collective dynamics in optomechanical arrays,"
%Phys. Rev. Lett {\bf 107}, 043603 (2011).

%\bibitem{EW}
%S.F. Edwards and D.R. Wilkinson,
%%``The surface statistics of a granular aggregate,"
%Proc. R. Soc. London, Ser. A {\bf 381}, 17 (1982).

%\bibitem{Gardiner_1985}
%C.W. Gardiner, {\it Handbook of Stochastic Methods} 2nd ed.
%(Springer-Verlag, New York, 1985).

%\bibitem{Goldenfeld_1992}
%N. Goldenfeld,
%{\it Lectures on Phase Transitions and the Renormalization Group}, (Addison-Wesley, New York, 1992).

%\bibitem{Monasson_EPJB1999}
%R. Monasson,
%%``Diffusion, localization and dispersion relations on `small-world' lattices,"
%Eur. Phys. J. B {\bf 12}, 555 (1999).

%\bibitem{Kozma_UGA2004}
%B. Kozma and G. Korniss,
%%``Stochastic growth in a small world,"
%in {\it Computer Simulation Studies in Condensed Matter Physics XVI}, edited by D.P. Landau, S.P. Lewis, and H.-B. Schttler,
%Springer Proceedings in Physics Vol. 95 (Springer-Verlag, Berlin, 2004) pp.~29-33.

%\bibitem{Kozma_PRL2004}
%B. Kozma, M. B. Hastings, and G. Korniss,
%%``Roughness scaling for Edwards-Wilkinson relaxation in small-world networks,"
%Phys. Rev. Lett. {\bf 92}, 108701 (2004).

%\bibitem{Kim_PRL2007}
%D.-H. Kim and A.E. Motter,
%%``Ensemble averageability in network spectra,"
%Phys. Rev. Lett. {\bf 98}, 248701 (2007).

%\bibitem{Bambi_JEDC2008}
%M. Bambi,
%%``Endogenous growth and time-to-build: the AK case,"
%J. Econ. Dynamics and Control {\bf 32}, 1015 (2008).

%\bibitem{Amann_PhysA2007}
%A. Amann, E. Scholl, and W. Just,
%%``Some basic remarks on eigenmode expansions of time-delay dynamics,"
%Physica A {\bf 373}, 191 (2007).

%\bibitem{vKampen_JSP1981}
%N.G. van Kampen,
%%``Ito versus Stratonovich,"
%J. Stat. Phys. {\bf 24}, 175 (1981).

%\bibitem{networkSpectra}
%A. E. Motter,
%%``Bounding network spectra for network design,"
%New J. Phys. {\bf 9} 182 (2007).

%\bibitem{Krantz1999}
%S.G. Krantz,
%{\it Handbook of Complex Variables}, (Birkhauser, Boston, MA, 1999).

%\bibitem{Hoefener2011}
%J.M. H\"ofener, G.C. Sethia, T. Gross,
%%``Stability and resonance in networks of delay-coupled delay oscillators,"
%Europhys. Lett. {\bf 95}, 40002 (2011).

%\bibitem{Luz_JCAM1996}
%T. Luzyanina and D. Roose,
%%``Numerical stability analysis and computation of Hopf bifurcation points for delay differential equations,"
%J. Comp. Appl. Math. {\bf 72}, 379 (1996).

%\bibitem{Atay_PRL2003}
%F.M. Atay,
%%''Distributed delays facilitate amplitude death of coupled oscillators,"
%Phys. Rev. Lett. {\bf 91}, 094101 (2003).

%\bibitem{Jadbabaie_IEEE2010}
%A. Papachristodoulou, A. Jadbabaie, and U. M\"unz,
%%``Effects of delay in multi-agent consensus and oscillator synchronization,"
%IEEE Trans. Automat. Contr. {\bf 55}, 1471 (2010).

%\bibitem{Raychaudhuri2001}
%S. Raychaudhuri, M. Cranson, C. Przybyla, and Y. Shapir
%%``Maximal height scaling of kinetically growing surfaces,"
%Phys. Rev. Lett. {\bf 87}, 136101 (2001).

\end{thebibliography}
\end{document}